\documentclass[
reprint, twocolumn, secnumarabic%
,tightenlines%
,amssymb,
amsmath,
nobibnotes,
nofootinbib, aps, prl, showpacs,showkeys]{revtex4}

\usepackage{subfigure}
\usepackage{tikz}
\usepackage{docs}%
\usepackage{bm}%
\usepackage{graphicx} %
\usepackage{rotating} %
\usepackage{hyperref} %
\usepackage{url} %
\usepackage{srcltx} %
\usepackage{float}
\expandafter \ifx \csname package@font\endcsname\relax\else
\expandafter \expandafter \expandafter
\usepackage
\expandafter \expandafter
\expandafter{\csname package@font\endcsname}%
\fi
\usepackage{epstopdf}
\usepackage{slashed}
\usepackage{ulem}
\usepackage{srcltx}
\usepackage{soul}
\usepackage{amsmath}
\usepackage{natbib} 
\usepackage{ulem} 

\usepackage{float} 
\extrafloats{200}
\pacs{13.15.+g, 13.60.Le}

\keywords{neutrino-nucleon interactions, single pion production, spin asymmetry, nucleon polarization, nonresonant background contribution}

\usepackage{epstopdf}

\usepackage{enumerate}

\begin{document}

\title{Spin asymmetry in single pion production induced by weak interactions of neutrinos with polarized nucleons}

\author{Krzysztof M. Graczyk}
\email{krzysztof.graczyk@uwr.edu.pl}

\author{Beata E. Kowal}

\affiliation{Institute of Theoretical Physics, University of Wroc\l aw, plac Maxa Borna 9,
50-204, Wroc\l aw, Poland}

\date{\today}%

\begin{abstract}
The single pion production (SPP) in the charged-current neutrino (antineutrino) scattering off the polarized nucleon is discussed. The spin asymmetry is predicted within two approaches. The spin
polarizations of the target nucleon that are  longitudinal and perpendicular to the neutrino momentum are considered. It is shown, in several examples, that information about the SPP dynamics	coming from the spin asymmetry  is complementary to information obtained from measurements of spin averaged cross section. Indeed, the spin asymmetry is sensitive to the nonresonance background description of the SPP model. For the normal polarization of the target, the spin asymmetry is given by the interference between the resonance and the nonresonance contributions. 

\end{abstract}

\maketitle

\section{Introduction}

The neutrino oscillation phenomenon has been investigated  for several decades. The oscillation parameters are relatively well established~\cite{wascko_morgan_2018_1286752}; however, two parameters,  $CP$-violation phase $\delta$ and the mixing angle  $\theta_{23}$, are still poorly known \cite{Valle:2018pgs}.

In the simplest two-flavor scenario, the probability for the oscillation $\nu_\alpha \to \nu_\beta$ reads 
$$P(\nu_\alpha \to \nu_\beta) \approx \sin^2(2\theta) \sin^2({\Delta m^2} L/{4 E}),$$ 
where $\Delta m$ is the neutrino mass difference, $\theta$ is a mixing angle,  $E$ is the neutrino energy, and
$L$ is the distance between  the source of the neutrinos and the detector. 

In the long baseline experiments, such as T2K \cite{wascko_morgan_2018_1286752} or Nova \cite{sanchez_mayly_2018_1286758}, the distance $L$ is known. The neutrino beam, produced at the accelerator, consists of mainly muon neutrinos of the energy of the order of 1~GeV.  However, the beam is not monochromatic and its energy profile is obtained from the analysis of the interaction of the neutrinos with the target. Therefore, the determination of the oscillation parameters depends on  the accuracy in estimation of the neutrino energy.  

Usually the neutrino energy is reconstructed from the analysis of the quasielastic (QE) neutrino-nucleus scattering. The reconstruction is based on the knowledge of the neutrino-nucleon and the neutrino-nucleus cross sections~\cite{doi:10.1146/annurev-nucl-102115-044720,sanchez_federico_2018_1295707}. However, in the $1$~GeV energy range a sizable fraction of the detected interactions is inelastic. In particular, the so-called  single pion production (SPP) processes are distinguished. The SPP events contribute to the background for the measurement of the QE scattering. Moreover, the neutral current $\pi^0$ production events can be wrongly identified as a   
the signal  for $\nu_\mu \to \nu_e$ oscillation.     

Intense studies of the fundamental neutrino properties have caused  new interest in the investigation of the neutrino-nucleon and the neutrino-nucleus scattering. In this work we focus on the problem of the single pion production in the neutrino-nucleon scattering in the energy range characteristic  of the long baseline neutrino oscillation experiments. This topic  has been  studied theoretically~\cite{Adler:1968tw,Rein:1980wg,Fogli:1979cz,Gershtein:1980vd,Rein:1987cb,Sato:2003rq,Hernandez:2006yg,Graczyk:2007bc,Leitner:2008ue,Graczyk:2009qm,Nakamura:2015rta,Serot:2012rd,Lalakulich:2010ss,Alam:2015gaa,Graczyk:2014dpa,Barbero:2008zza,Alvarez-Ruso:2015eva,Gonzalez-Jimenez:2016qqq,Hernandez:2016yfb,Yao:2018pzc,Yao:2019avf} and experimentally \cite{Radecky:1981fn,Kitagaki:1986ct,Rodriguez:2008aa,AguilarArevalo:2010bm,McGivern:2016bwh,Abe:2016aoo} for the last 50 years.   

The SPP scattering amplitude is dominated by the resonance (RES) contribution given by a weak nucleon-resonance transition. However, a complete SPP model should include also the diagrams describing the so-called nonresonance background (NB) terms. The way the RES and NB contributions are treated gives rise to the differences between various theoretical approaches.  

In order to test the SPP models their predictions must be confronted  with the experimental measurements of the neutrino-nucleon and the neutrino-nucleus cross sections.  As we explained  in our previous paper  \cite{Graczyk:2017rti}, the spin averaged cross sections contain only a part of the information about the dynamical structure of the SPP amplitudes. Complementary information  can be obtained from the analysis of the  polarization transfer (PT) observables. 

The investigation of the PT in the neutrino-nucleon and the neutrino-nucleus scattering has been discussed since the 1960s \cite{Adler1963,Pais:1971er,LlewellynSmith:1971uhs,Kuzmin:2003ji,Hagiwara:2003di,Graczyk:2004uy,Kuzmin:2004yb,Bilenky:2013iua,Bilenky:2013fra,Akbar:2016awk,Akbar:2017qsf,Fatima:2018gjy,Fatima:2018tzs}.
Recently in \cite{Graczyk:2017rti,Graczyk:2017ngi}, we reported the results of the discussion of  the impact of the NB contribution on the PT observables. It was shown that  the  components of the polarizations  of the charged lepton and the final nucleon contain  unique information about the relative phase between the RES and NB amplitudes which can be used to  constrain theoretical models, in particular, the description of the nonresonant background.   
\begin{figure}[tb!]
		\includegraphics[width=0.5\textwidth]{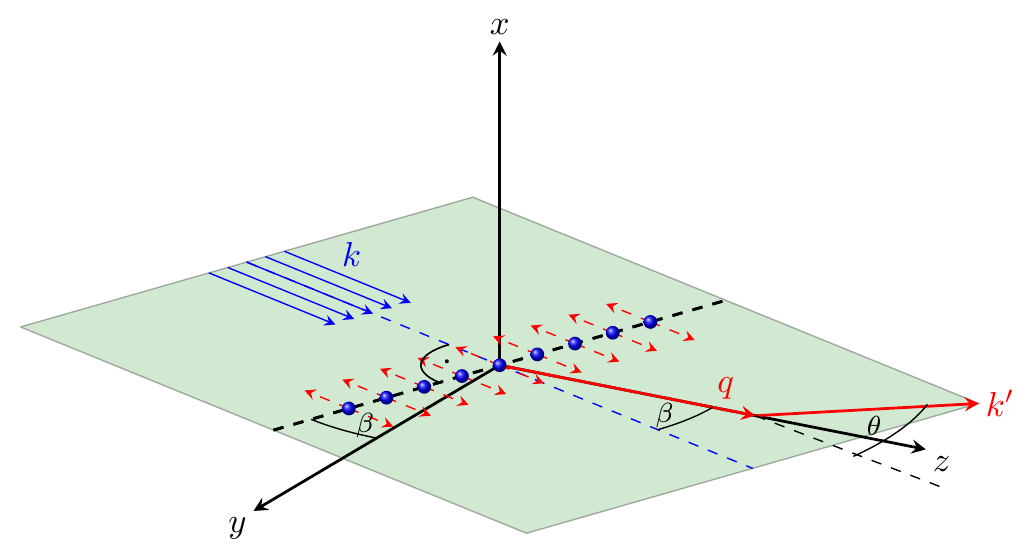}
\caption{The longitudinal target  polarization. The full balls denote the target; the direction of the polarization is indicated by the dashed arrow. \label{Fig_polarization_V1}}
\end{figure}
\begin{figure}
		\includegraphics[width=0.5\textwidth]{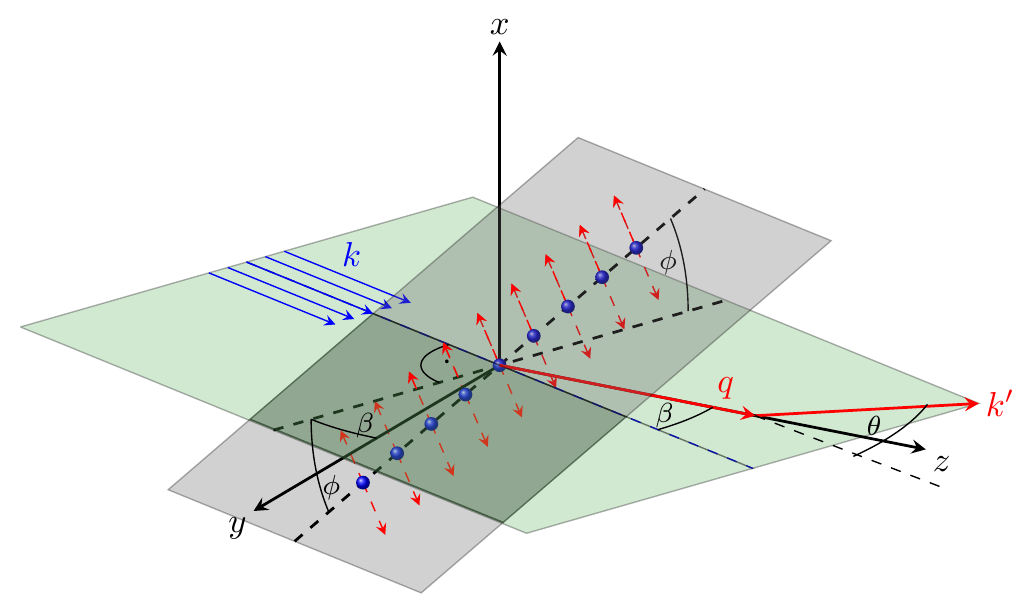}
	\caption{The perpendicular target polarization.  The full balls denote the target; the direction of the polarization is indicated by the dashed arrow. \label{Fig_polarization_V2}}
\end{figure}

In this report,  instead of analyzing the polarizations of the final particles,  the  neutrino scattering off the polarized target is considered. We propose to investigate properties of a spin asymmetry observable.  A similar quantity was discussed for the elastic electron-nucleon and the electron-nucleus scattering~\cite{Dombey:1969wk,Donnelly:1985ry}. Indeed, the measurement of the asymmetry in the electron-nucleon scattering  was proposed as an alternative technique to the Rosenbluth method for obtaining the electric and magnetic form factors of the nucleon. \footnote{The first measurements of the spin asymmetry are reported in Ref.~\cite{Alguard:1976bk}.} In this work we calculate and analyze the spin asymmetry in the  SPP  induced by interactions of the  neutrinos with the nucleons. We show that this observable  is sensitive to the NB contribution.  Hence,   the spin asymmetry contains  unique information about the SPP dynamics not accessible in the spin averaged cross section measurements. 

Similarly as in \cite{Graczyk:2017rti}, two different SPP approaches are considered \cite{Hernandez:2007qq,Fogli:1979cz}. Our studies are restricted to the neutrinos of the energy of the order of 1~GeV. Therefore, to model the RES contribution,  we consider only the weak $N\to \Delta(1232)$ transition.     The predictions are made for full models (RES and NB contributions) and the version of the models with resonance contribution only.   

The paper is organized as it follows. In Sec.~\ref{Sec_Spin_asymmetry} the necessary formalism is introduced, Sec.~\ref{Sec_Results} presents the numerical results and their discussion, and a summary is given in Sec.~\ref{Sec_summary}.

\section{Spin asymmetry}

\label{Sec_Spin_asymmetry}
Let us consider the SPP processes induced by the charged current  muon neutrino (antineutrino) interactions with the polarized nucleon target, namely,
\begin{eqnarray}
\label{Process_polarized_target_neutrino}
\nu_\mu(k) + \vec{N}(p,\mathbf{s}) & \to &  \mu^{-}(k') + N'(p')+\pi(k_{\pi}),\\
\label{Process_polarized_target_antyneutrino}
\overline{\nu}_\mu(k) + \vec{N}(p,\mathbf{s}) & \to &  \mu^{+}(k') + N'(p')+\pi(k_{\pi}),
\end{eqnarray}
where $k^\alpha=(E,{\bf k})$ and ${k'}^\alpha=(E',{\bf k'})$ are the four-momenta of the initial and the final leptons, respectively, while $p^\alpha=(E_p,{\bf p})$; ${p'}^\alpha=(E_{p'},{\bf p'})$; and $k_\pi^\alpha=(E_{\pi},{\bf k_{\pi} })$ denote the four-momenta of the incoming nucleon (N), the outgoing nucleon ($N'$), and the pion, respectively. The calculations are made in the laboratory frame; hence, the spin four-vector of the target reads  
\begin{equation}
s^\mu = (0,\mathbf{s}),
\end{equation}
where $\mathbf{s}^2=1$. 

The four-momentum transfer is denoted by
\begin{equation}
q^\alpha \equiv k^\alpha-{k'}^\alpha = (\omega,\mathbf{q}), 
\end{equation}
where $\omega$ and $\mathbf{q}$ denote the transfer of the energy and the momentum, respectively. 

Let us introduce the hadronic invariant  mass 
\begin{equation}
W \equiv (p+q)^2
\end{equation} and 
\begin{equation}
Q^2 \equiv -q^2.
\end{equation}  
Eventually, let $\Omega(\theta,\phi)$ denote  a solid angle depending on $\theta \equiv \angle(\mathbf{k},\mathbf{k'})$  and  $\phi $ is a corresponding  azimuth angle.


We define a spin asymmetry  by the ratio
\begin{equation}
\label{Eq::Assymetry}
\mathcal{A}(\mathbf{s},d\sigma) = \frac{d \sigma (\mathbf{s}) - d \sigma (-\mathbf{s})}{d \sigma (\mathbf{s}) + d \sigma (-\mathbf{s})},
\end{equation}
where $d\sigma $ is the differential cross section. The asymmetry is linear in $\mathbf{s}$, namely,
$    \mathcal{A} = \mathbf{s} \cdot \mathbf{a}$.
\begin{figure*}[t]
\includegraphics[width=\textwidth]{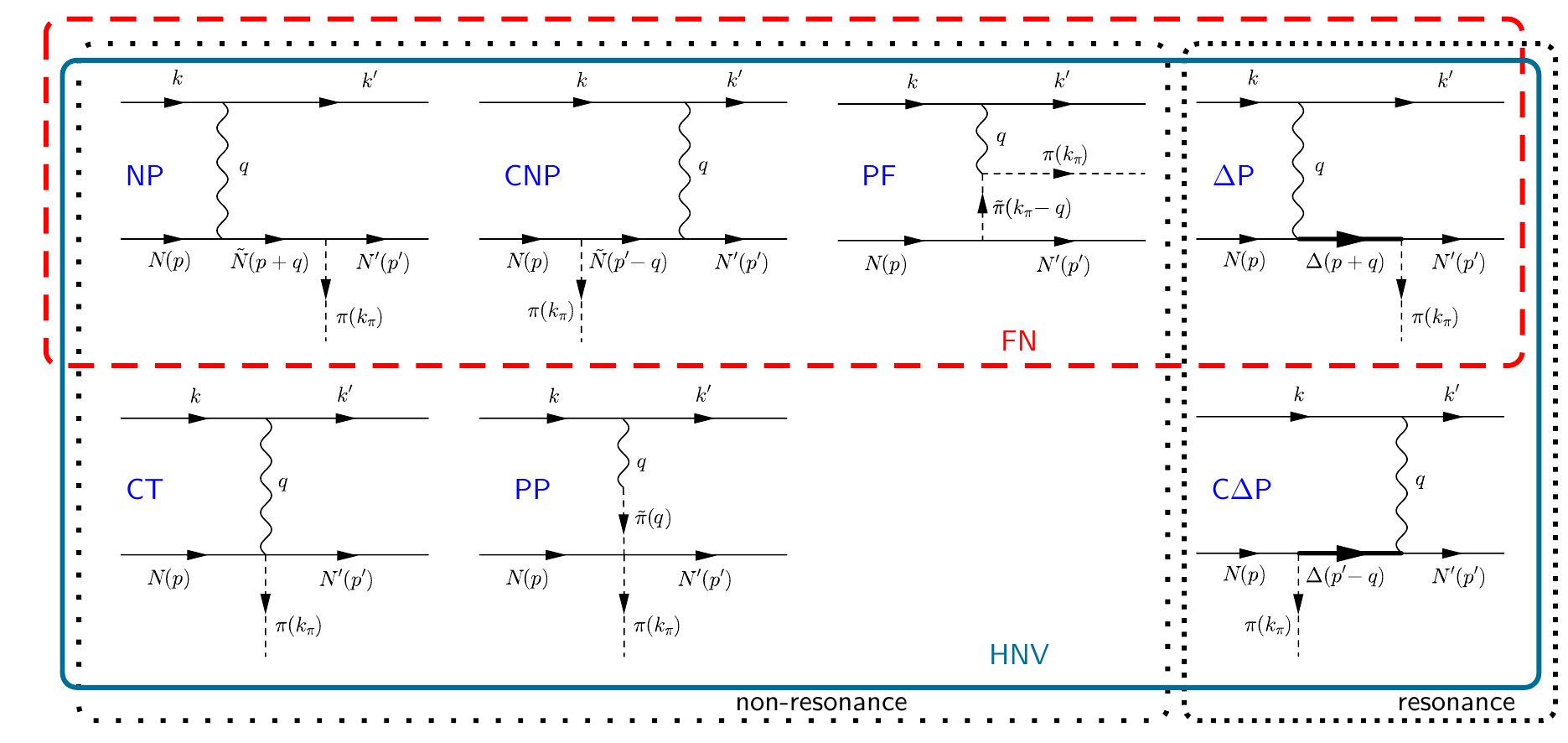}
\caption{Diagrams contributing to the SPP induced by $\nu N$ interaction in the HNV (blue solid frame) and the FN (red dashed frame) models. The resonance contribution is framed by the densely dotted line. The nonresonance contribution is framed by the loosely dotted line. NP denotes the nucleon pole, CNP the conjugate nucleon pole, CT the contact term, PP the pion pole, PF the pion in flight, $\Delta P$ the $\Delta$ pole, and $C\Delta P$ the conjugate $\Delta$ pole. \label{Fig_model_diagrams}}
\end{figure*}
\newpage
\begin{figure*}[t]
\includegraphics[width=0.8\textwidth]{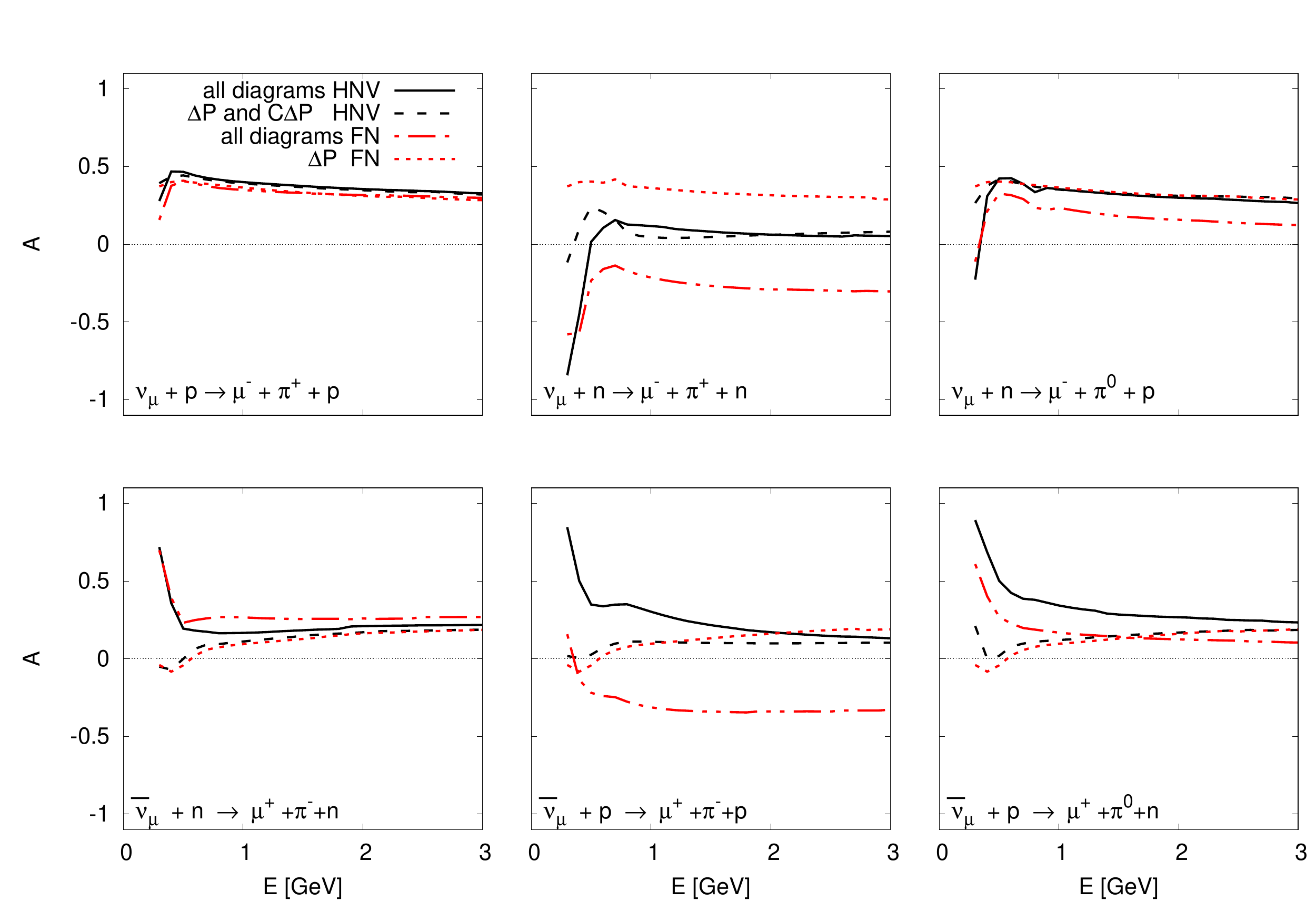}
	\caption{  Dependence of the asymmetry $\mathcal{A}_\parallel(E,\sigma)$ on the energy.  The solid (dashed) line denotes the full (RES) model contributions of the HNV model while dotted-dashed (dotted) line denotes 
	 the full (RES) model contributions of the FN  model.  The resonance contribution is given by $|\mathcal{M}_{\Delta P} + \mathcal{M}_{C\Delta P}|^2$ and $|\mathcal{M}_{\Delta P}|^2$ in the HNV and the FN models, respectively.  \label{Fig_asymmetry_sigma_longitudinal}}
\end{figure*}
\begin{figure*}[tb!]
	\includegraphics[width=0.9\textwidth]{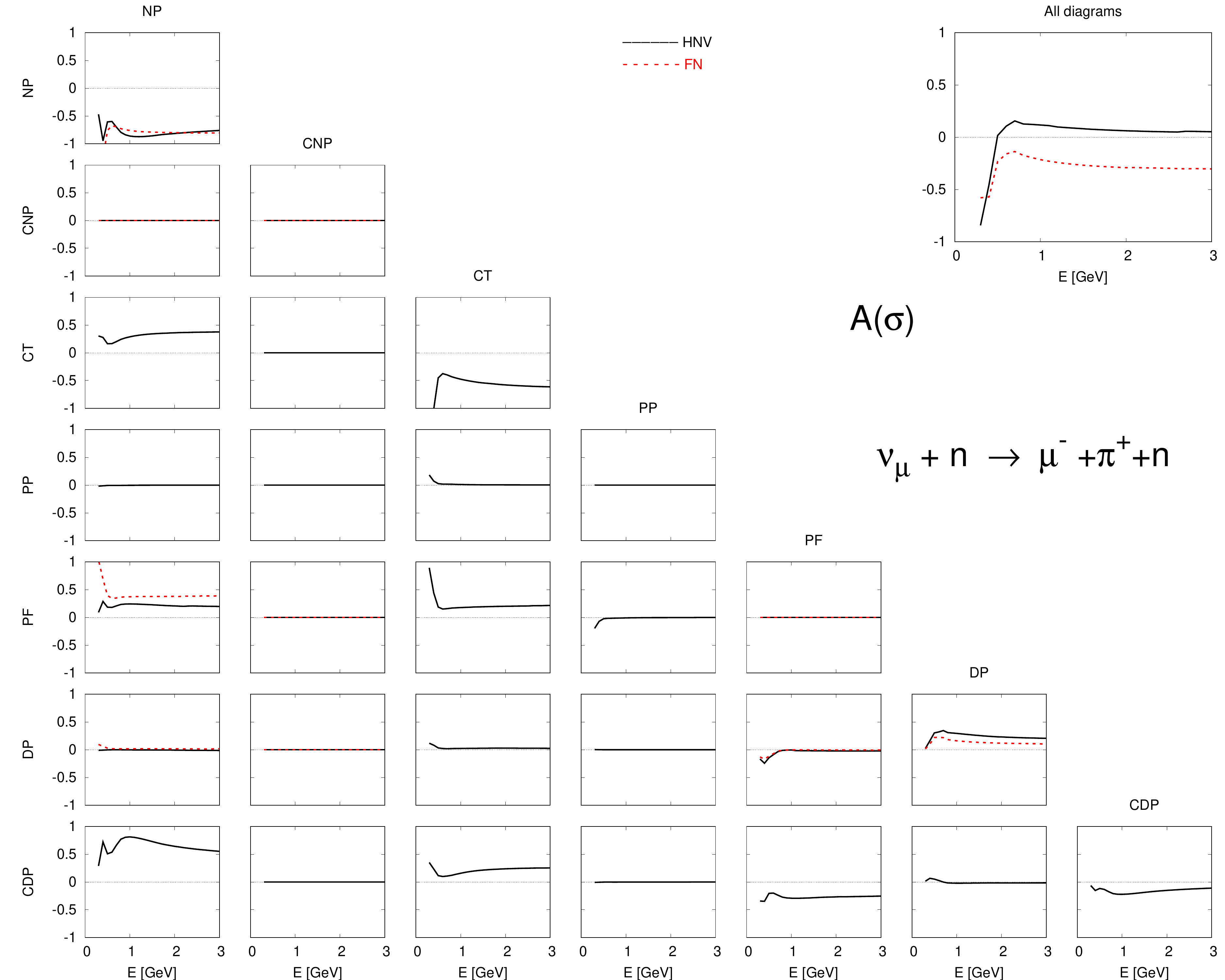}	
	\caption{
	Decomposition of the longitudinal asymmetry  $\mathcal{A}_\parallel(E,\sigma)$, for process \eqref{C1}, into contributions from interference between diagrams. On the diagonal, the contribution from $|\mathcal{M}_d|^2$ ($\mathcal{M}_d$ denotes the matrix element of a $d$-diagram) is given, while below the diagonal, the interference terms  $2\Re (\mathcal{M}_i \mathcal{M}_j^*)$ ($j$ denotes the column and $i$ the row) are plotted. The solid black (dotted red) line represents the HNV (FN) model predictions.
   \label{Fig_asymmetry_SigmaChan1}}
\end{figure*}
\newpage

Two variants of the nucleon polarization are studied, namely,
\begin{enumerate}[(i)]
	\item \label{variant_1} Longitudinal polarization:    Nucleon polarized along the momentum of the incoming neutrino (see Fig.~\ref{Fig_polarization_V1}). In this case, $\mathbf{s} = \mathbf{s}_\parallel$ and 
	\begin{equation}
	\label{Eq_Aparallel}
	   \mathcal{A}(\mathbf{s}_\parallel,d\sigma) \equiv \mathcal{A}_\parallel(d \sigma)  ;
	\end{equation}
	
	\item \label{variant_2} Perpendicular polarization: Nucleon polarized along an $x$ axis which is perpendicular to the neutrino momentum (Fig. \ref{Fig_polarization_V2}). In this case, $\mathbf{s} = \mathbf{s}_\perp$ and 
	\begin{equation}
	\label{Aperpendicular}
	   \mathcal{A}(\mathbf{s}_\perp,d\sigma) \equiv \mathcal{A}_\perp(d\sigma)  .
	\end{equation}     
\end{enumerate}
Notice that in the last variant the $\phi$-dependence of $d\sigma /d \Omega$  cannot be trivially integrated out. Indeed the rotational symmetry (along $\mathbf{k}$) is broken by the choice of the direction of the target's spin. In order to perform calculations, we choose the coordinates so that 
\begin{equation}
\phi = \angle(\mathbf{s}_\perp,\hat{\mathbf{n}}),
\end{equation} 
where $\hat{\mathbf{n}}$ is the normal vector of the scattering plane spanned by $\mathbf{k}$ and $\mathbf{k'}$.
%

\newpage
\begin{figure*}[tb!]
\includegraphics[width=0.8\textwidth]{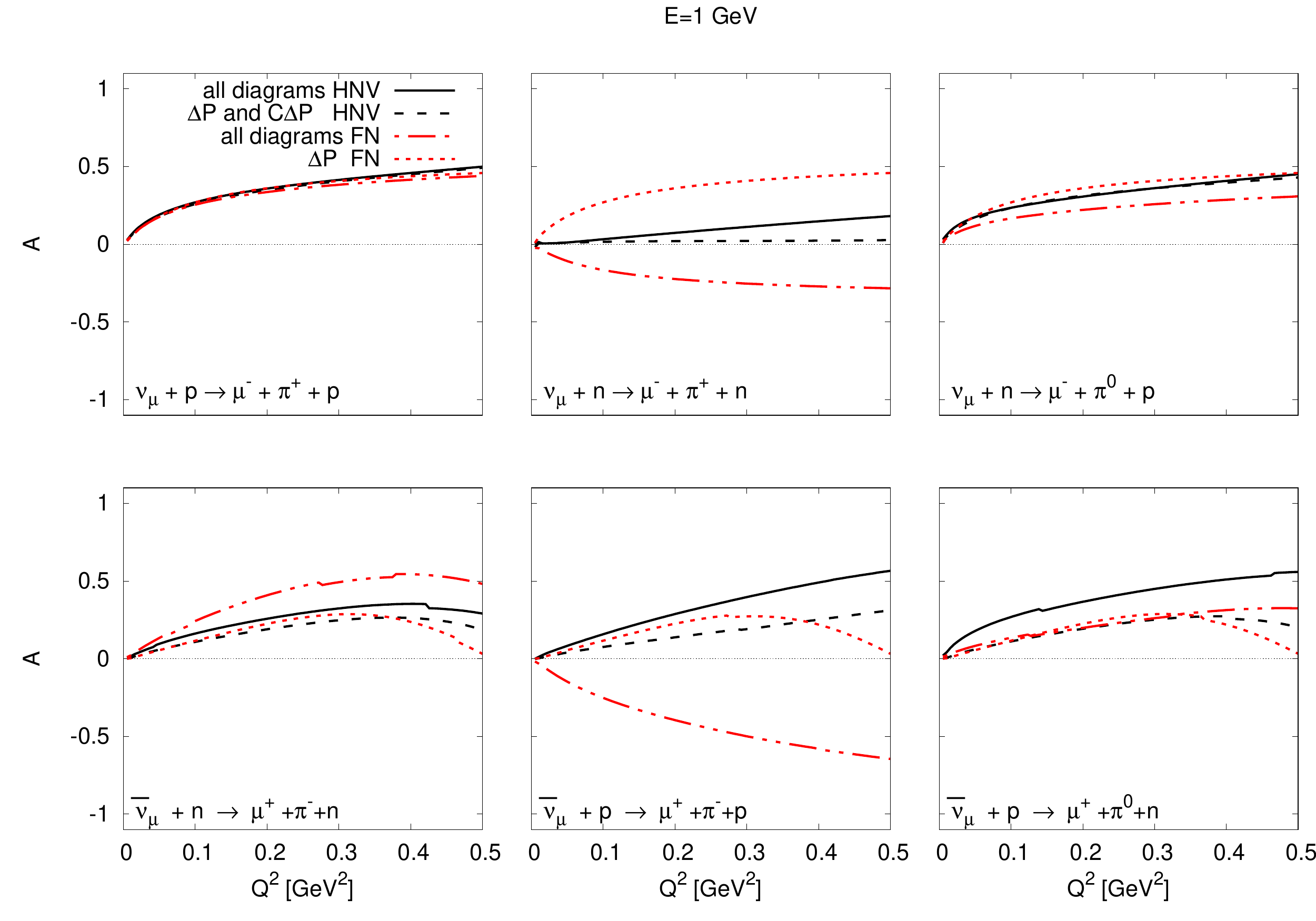} 
	\caption{  Dependence of the asymmetry $\mathcal{A}_\parallel(E=1~\mathrm{GeV},d\sigma/dQ^2)$ on $Q^2$.  The solid (dashed) line denotes the full (RES) model contributions of the HNV model, while the dotted-dashed (dotted) line denotes 
		 the full (RES) model contributions of the FN  model. The RES contribution is given by $|\mathcal{M}_{\Delta P} + \mathcal{M}_{C\Delta P}|^2$ and $|\mathcal{M}_{\Delta P}|^2$ in the HNV and the FN models, respectively.  \label{Fig_asymmetry_sigma_longitudinal_E1_Q2}}
\end{figure*}
\begin{figure*}[t]
	\includegraphics[width=0.8\textwidth]{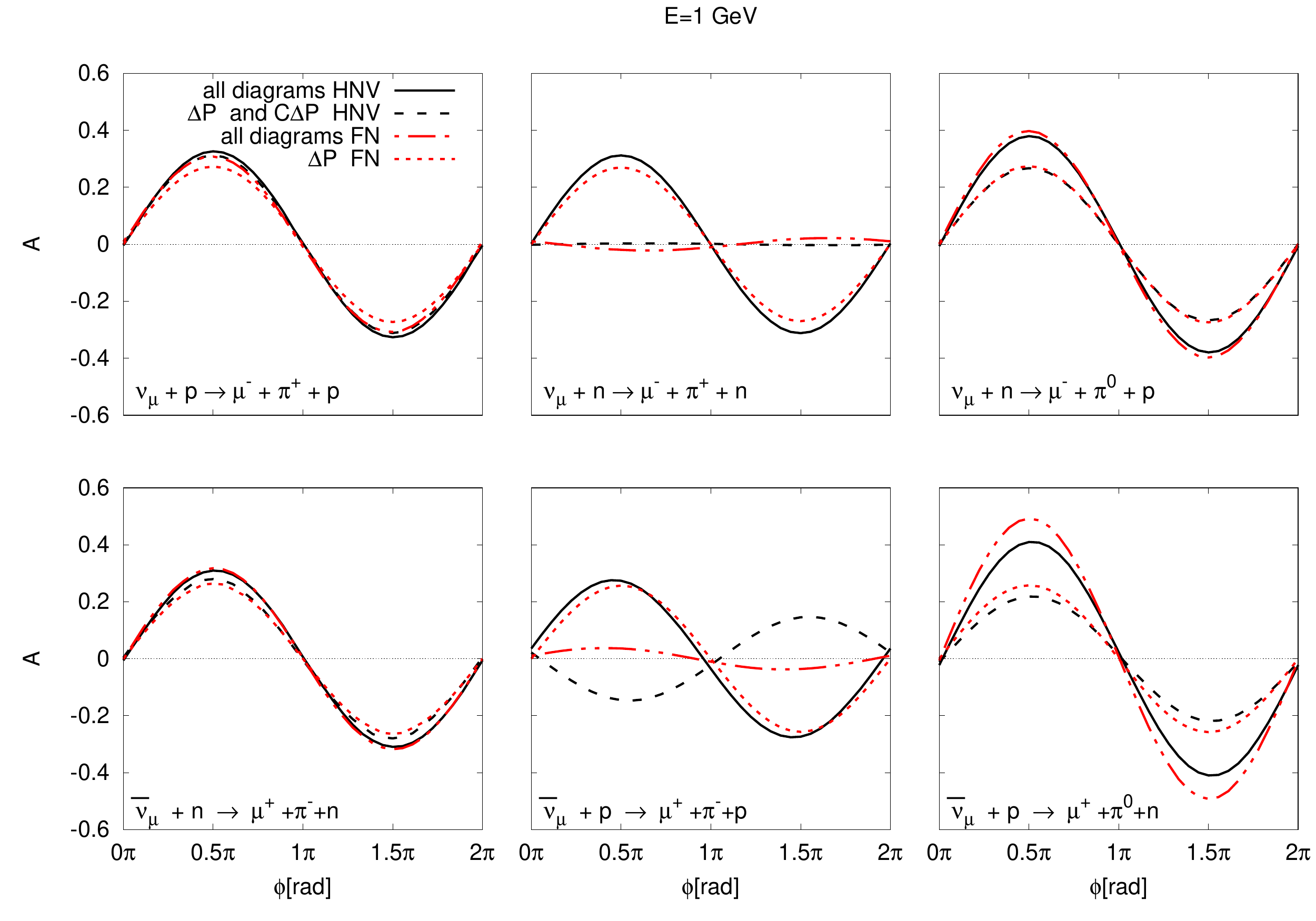}
	\caption{Dependence of the asymmetry $\mathcal{A}_\perp(E=1~\mathrm{GeV},d\sigma/d\phi)$ on the $\phi$ angle.  The solid (dashed) line denotes the full (RES) model contributions of the HNV model, while the dotted-dashed (dotted) line denotes 
		 the full (RES) model contributions of the FN  model.  The RES contribution is given by $|\mathcal{M}_{\Delta P} + \mathcal{M}_{C\Delta P}|^2$ and $|\mathcal{M}_{\Delta P}|^2$ in the HNV and the FN models, respectively. \label{Fig_asymmetry_E1GeV_phi_dependence}}
\end{figure*}
\begin{figure*}[b]	
\includegraphics[width=0.8\textwidth]{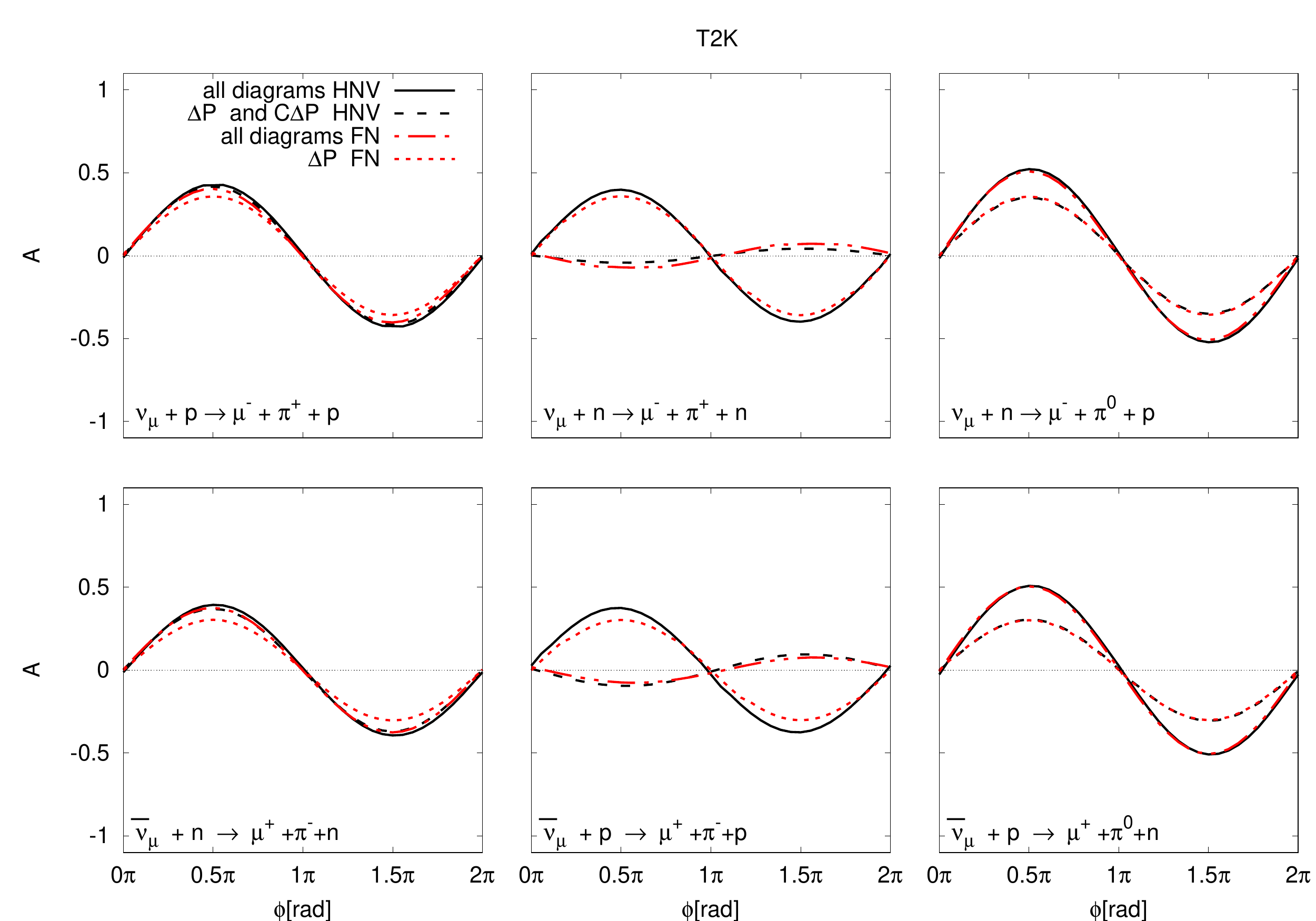}
	\caption{	 The dependence of the asymmetry $\mathcal{A}_\perp(\theta=10^{\circ},d^2\sigma/d\Omega)$ [see Eq. (\ref{As_flux})]  on the $\phi$ angle calculated for a T2K energy distribution.   The solid (dashed) line denotes the full (RES) model contributions of the HNV model, while the dotted-dashed (dotted) line denotes 
		 the full (RES) model contributions of the FN  model.  The RES contribution is given by $|\mathcal{M}_{\Delta P} + \mathcal{M}_{C\Delta P}|^2$ and $|\mathcal{M}_{\Delta P}|^2$ in the HNV and the FN models, respectively. \label{Fig_asymmetry_fluxT2K} }
\end{figure*}
\newpage
\begin{figure*}[bt!]
	\includegraphics[width=0.9\textwidth]{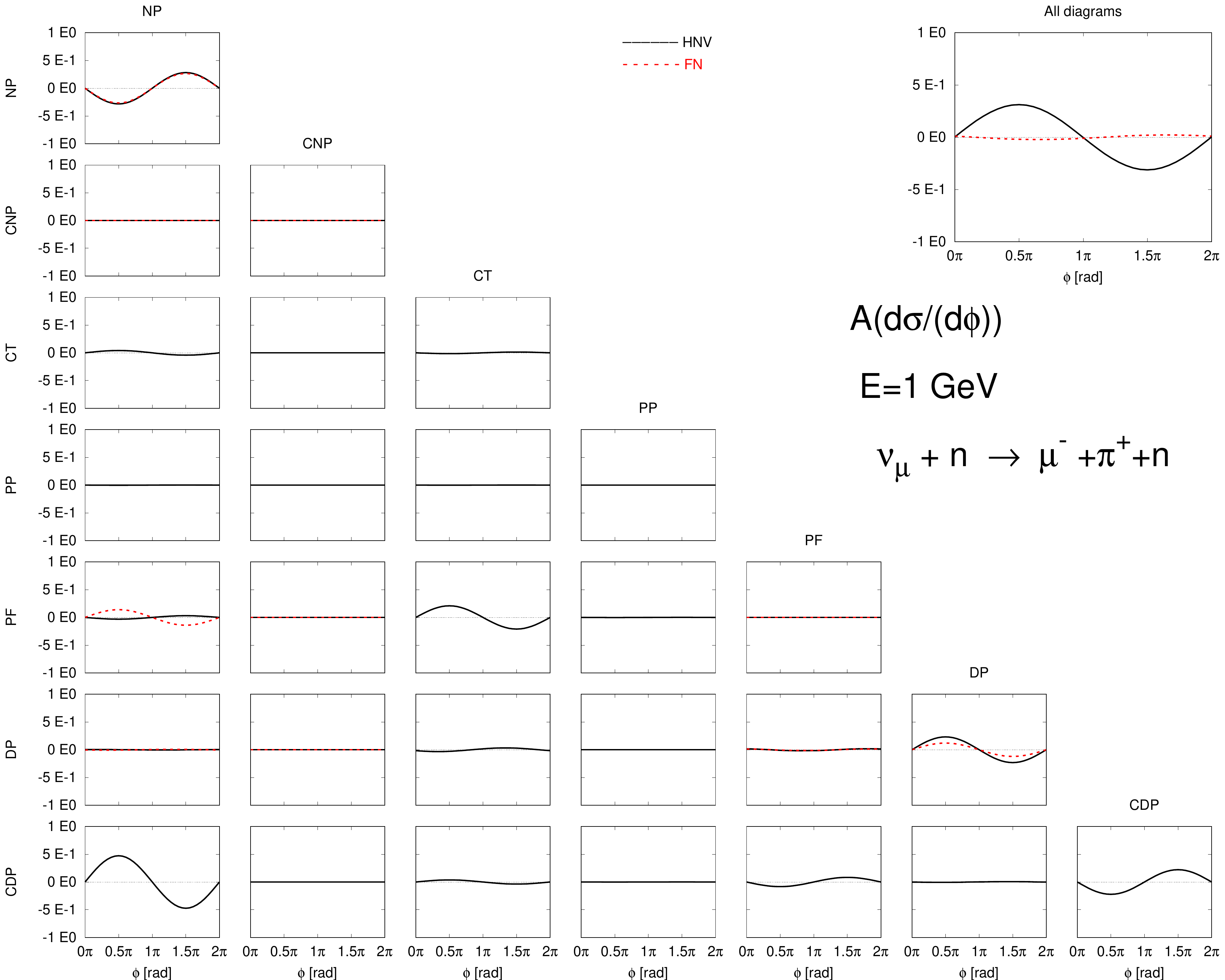}
	\caption{	Decomposition of $\mathcal{A}_\perp(E=1~\mathrm{GeV},d\sigma/d\phi)$, for the process \eqref{C1}, into contributions from interference between diagrams. On the diagonal, the contribution from $|\mathcal{M}_d|^2$ ($\mathcal{M}_d$ denotes the matrix element for a $d$-diagram) is given, while below the diagonal, the interference terms  $2\Re (\mathcal{M}_i \mathcal{M}_j^*)$ ($j$ denotes the column and $i$ the row) are plotted. The solid black (dotted red) line represents the HNV (FN) model predictions.\label{Fig_asymmetry_E1GeVchan1}}
\end{figure*}
\begin{figure*}[bt!]
	\includegraphics[width=0.85\textwidth]{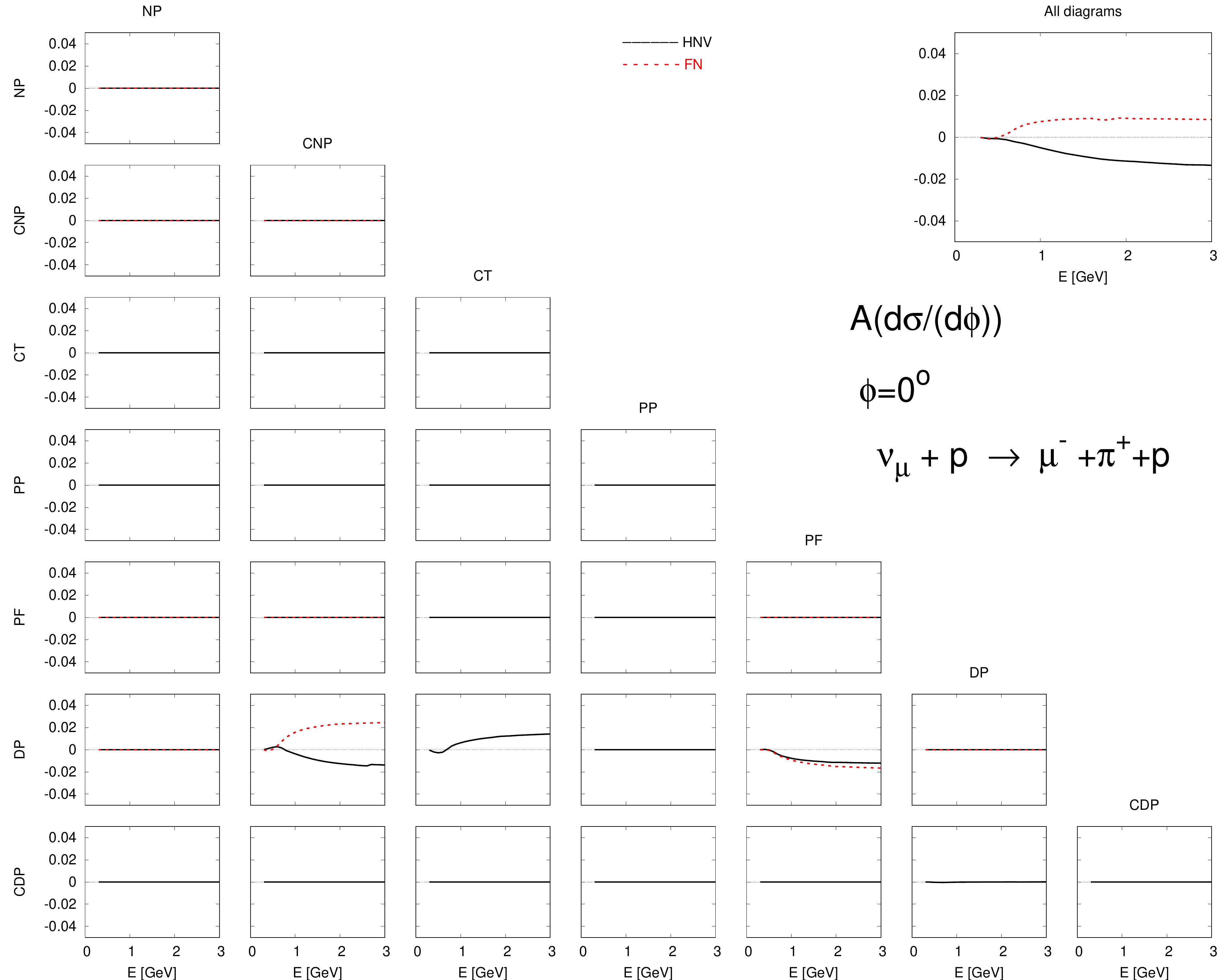}
	\caption{Decomposition of the perpendicular asymmetry  $\mathcal{A}_\perp(\phi=0^o,d\sigma/d\phi)$, for the process \eqref{C0}, into contributions from interference between diagrams. On the diagonal, the contribution from $|\mathcal{M}_d|^2$ ($\mathcal{M}_d$ denotes the matrix element for a $d$-diagram) is given, while below the diagonal, the interference terms  $2\Re (\mathcal{M}_i \mathcal{M}_j^*)$ ($j$ denotes column and $i$ the row) are plotted. The solid black (dotted red) line represents the HNV (FN) model predictions.
    \label{Fig_asymmetry_Phi0chan0}}
\end{figure*}
\begin{figure*}[bt!]
	\includegraphics[width=0.8\textwidth]{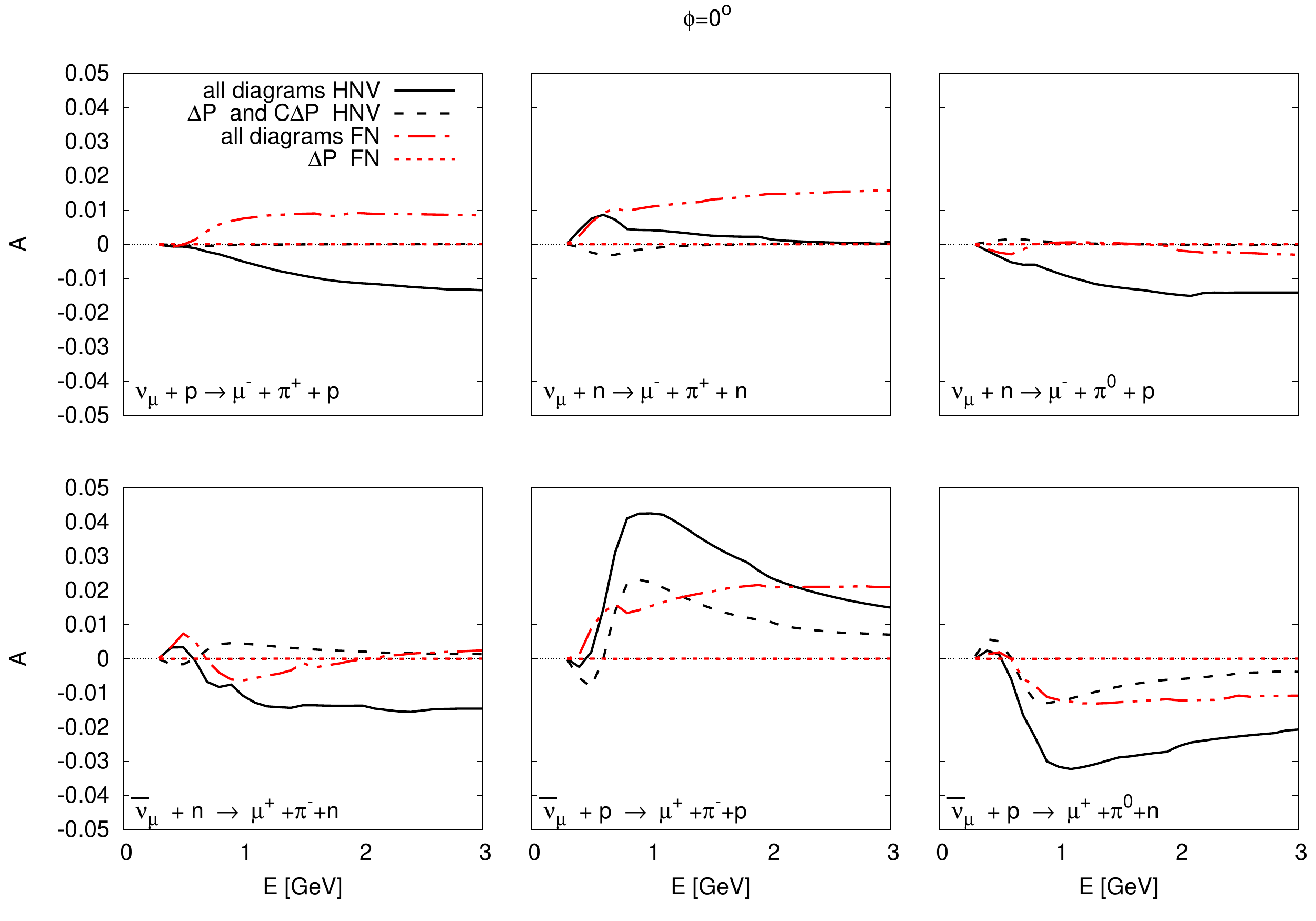} 
	\includegraphics[width=0.8\textwidth]{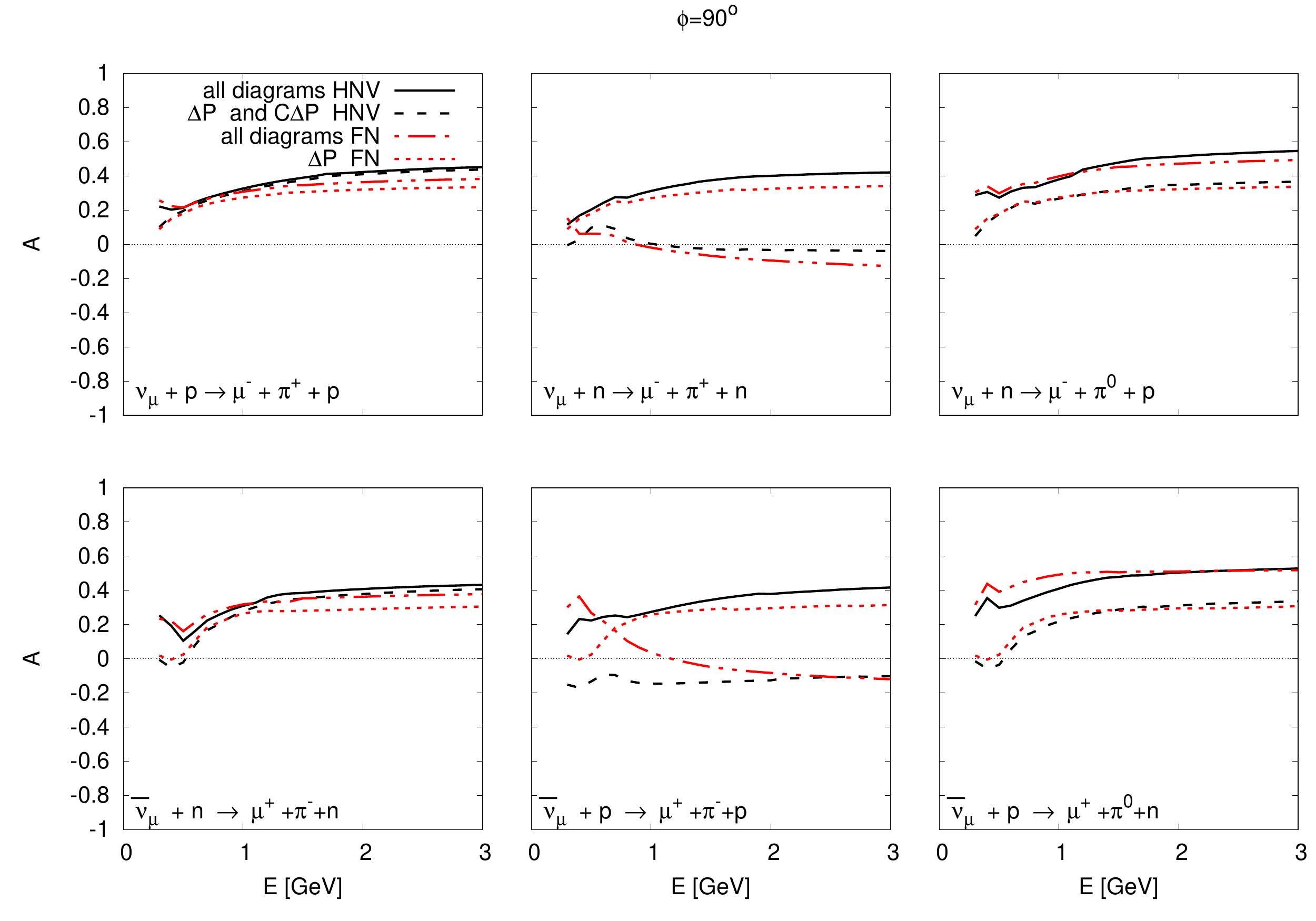} 
	\caption{	Dependence of the asymmetry $\mathcal{A}_\perp(d\sigma/d\phi)$ on energy calculated for $\phi=0^{\circ}$ (top panel) and $\phi=90^{\circ}$ (bottom panel).  The solid (dashed) line denotes the full (RES) model contributions of the HNV model while dotted-dashed (dotted) line denotes 
		 the full (RES) model contributions of the FN  model.  The RES contribution is given by $|\mathcal{M}_{\Delta P} + \mathcal{M}_{C\Delta P}|^2$ and $|\mathcal{M}_{\Delta P}|^2$ in the HNV and the FN models, respectively.
 	 \label{Fig_asymmetry_phi0=0}}
\end{figure*}

\section{Results and discussion}
\label{Sec_Results}

\subsection{Numerical implementation}

Our main objective is to study  the properties of the spin asymmetry, in particular, its sensitivity to the NB contribution.  To  achieve this goal, similarly as in our previous work \cite{Graczyk:2017rti}, in order to perform the calculations, two SPP approaches are considered: the model by Hernandez, Nieves, and Valverde (HNV), as described in \cite{Hernandez:2007qq}, and the model by  Fogli and Nardulli (FN), as given in \cite{Fogli:1979cz}. In both descriptions, the scattering amplitude is calculated in a tree level approximation.

The predictions of the spin asymmetry for neutrino (antineutrino) scattering off a longitudinally  (\ref{Eq_Aparallel}) and perpendicularly (\ref{Aperpendicular}) polarized target are made for six  charged-current SPP processes:
 \begin{eqnarray}
 \label{C0}   \nu_\mu + \vec{p}  & \to & \mu^- + p + \pi^+ 
 \\
 \label{C1} \nu_\mu + \vec{n}  & \to & \mu^- + n + \pi^+
 \\
 \label{C2}  \nu_\mu + \vec{n} & \to & \mu^- + p + \pi^0
 \\
\label{C3}  \bar{\nu}_{\mu} + \vec{n} & \to & \mu^+ + n +\pi^-
\\
 \label{C4}  \bar{\nu}_{\mu} + \vec{p} & \to & \mu^+ + p +\pi^- 
\\
 \label{C5}    \bar{\nu}_{\mu} + \vec{p} & \to & \mu^+ + n +\pi^0. 
\end{eqnarray}

The differential cross section, for  the given SPP  processes, has the structure
\begin{equation}
    \frac{d^9 \sigma }{d \Omega d E' d^3 p' d^3 k_\pi} \sim \left| \sum_{a\in D}c_a \mathcal{M}_a\right|^2 \delta^{(4)}(p+k - k'-p'-k_\pi),
\end{equation}
where $D$ is the set of the diagrams, $c_a$ is the Clebsch-Gordan coefficient, and $\mathcal{M}_a$ is a matrix element for a diagram $a$. 

The full amplitude of the HNV model consists of contributions from seven diagrams. The NB amplitudes are  obtained from the nonlinear sigma model. The SPP contribution in the FN approach is given by five diagrams, where the NB contribution is motivated by the linear sigma model. All diagrams are plotted in Fig.~\ref{Fig_model_diagrams}. Our discussion is restricted to the first resonance region; hence, all calculations are performed for $W<1.4$~GeV. 

In the HNV model the NB contribution is given by the following diagrams,  nucleon-pole (NP), conjugate nucleon-pole (CNP), contact term (CT), pion in flight (PF) and pion-pole (PP).  The $\Delta(1232)$ resonance contribution is described by two diagrams: $\Delta P$ denotes the delta pole, and $C\Delta P$ the conjugate delta pole. 

The NB contribution in the FN model  consists of  three diagrams: pion in flight ($PF$) and  two  nucleon pole diagrams, $NP$ and $CNP$. But in the latter two diagrams the pseudoscalar pion-nucleon coupling is implemented, in contrast to the HNV model, where the pseudovector coupling is considered. The weak $N \to \Delta(1232)$ transition is oversimplified. Indeed, there is only one resonance diagram and the $ N W^- \Delta$ vertex is described by only two form factors.  

More details about the implementation of both models, the choice of the transition form factors etc., can be found in our previous paper~\cite{Graczyk:2017rti}.

\subsection{Spin asymmetry for longitudinal polarized target}

Figure~\ref{Fig_asymmetry_sigma_longitudinal}  presents the plots of the longitudinal spin asymmetry
$\mathcal{A}_\parallel(\sigma,E)$
calculated for the neutrino and the antineutrino scattering off the polarized target.  The  asymmetry varies  from $-0.5$ to $0.5$. Above $E= 1$~GeV, $\mathcal{A}_\parallel(\sigma,E)$ weakly depends on the neutrino energy. 

For the channels  (\ref{C0}) and (\ref{C3}) (related by the isospin symmetry), the NB contribution to $\mathcal{A}_\parallel(\sigma,E)$ is negligible. Indeed, in this case  the resonance contribution, from $\Delta P$, is dominant. Therefore,  
the predictions of $\mathcal{A}_\parallel(\sigma,E)$ within the HNV and the FN models are very similar. 
However, for the other channels, the asymmetry is quite model dependent.    Indeed, for the  processes \eqref{C1} and  \eqref{C4} (also related by the isospin symmetry), the asymmetries predicted within the HNV and FN models have completely different functional dependence (different sign and magnitude).    Moreover, in this case the NB contribution  is large and modifies  significantly $\mathcal{A}_\parallel(\sigma,E)$. To understand this property, we present Fig.~\ref{Fig_asymmetry_SigmaChan1}, where the contributions to the spin asymmetry from various diagrams are distinguished. It can be noticed that the deviations between  the HNV and FN models are due to the presence of the diagrams $C\Delta P$ and $CT$ in the HNV model. Eventually,  in both approaches the diagram  $NP$ gives rise  to the  difference between the full and the RES   model predictions.   

Similar observations, as above, can be made when $\mathcal{A}_\parallel(d\sigma/dQ^2,E)$ is examined; see Fig.~\ref{Fig_asymmetry_sigma_longitudinal_E1_Q2}.

\subsection{Spin asymmetry for perpendicularly  polarized target}

The spin asymmetry is given by the scalar product $\mathbf{s}\cdot\mathbf{a}$. In the case of the perpendicularly polarized target, the components of $\mathbf{s}_\perp$ are proportional to either $\sin(\phi)$ or $\cos(\phi)$. As the result, the spin asymmetry can be written in the form
\begin{equation}
\label{A_angular_decomposition}
	\mathcal{A}_\perp(\phi) = a_1 \cos(\phi)  + a_2  \sin(\phi).
\end{equation} 

The $\mathcal{A}_\perp(\phi)$ is dominated by  the sinusoidal part. It is shown in  Fig.~\ref{Fig_asymmetry_E1GeV_phi_dependence},  where the asymmetry $\mathcal{A}_\perp(d\sigma/d\phi)$ is plotted. The sinusoidal character is maintained also when  the asymmetry is calculated for the flux averaged cross sections, as it is illustrated in Fig.~\ref{Fig_asymmetry_fluxT2K}, where the $\phi$-dependence of 
\begin{equation}
\label{As_flux}
 \mathcal{A}_\perp(\phi,\theta=10^o) = \frac{ \displaystyle \int d E \Phi(E) \left[ \frac{d^2 \sigma(\mathbf{s}_\perp)}{d \Omega} - \frac{d^2 \sigma(-\mathbf{s}_\perp)}{d \Omega}\right]}{\displaystyle \int d E \Phi(E) \left[ \frac{d^2 \sigma}{d \Omega}(\mathbf{s}_\perp) + \frac{d^2 \sigma}{d \Omega}(-\mathbf{s}_\perp)\right]}
\end{equation}
calculated for the energy spectrum, $\Phi$,  of the T2K experiment~\cite{Abe:2012av}  is plotted. 

Similarly as in the case of the longitudinally polarized target for two channels  (\ref{C0}) and (\ref{C3}), the asymmetry is dominated by the resonance contribution of the $\Delta P$ diagram.  Hence, in this case, $\mathcal{A}_\perp(\phi)$ is insensitive to details of the NB model. 

It is important to remember that the FN model does not contain the $C\Delta P$ diagram, required by gauge invariance. Lack of this contribution leads to the  deviation between predictions of the   $\mathcal{A}_\perp(\phi)$ obtained for the FN and the HNV models for the channels (\ref{C1}) and (\ref{C4}); see Figs. \ref{Fig_asymmetry_E1GeV_phi_dependence} and \ref{Fig_asymmetry_fluxT2K} as well as Fig.~\ref{Fig_asymmetry_E1GeVchan1}. In the latter figure the decomposition of the asymmetry into contributions from various diagrams is shown.  

The spin asymmetry (\ref{A_angular_decomposition}) has contributions from $a_2 \sin(\phi) $ and $a_1 \cos(\phi)$; however, plots of Figs. \ref{Fig_asymmetry_E1GeV_phi_dependence} and  \ref{Fig_asymmetry_fluxT2K} suggest that the $a_2$ is dominant. It is interesting to remark that the $a_1$ component, connected with cosine, is given by the interference  between the resonance and nonresonance amplitudes.\footnote{There is also a small but  non-negligible contribution from the interference between $\Delta P$  and $C\Delta P$ diagrams.} Hence, any deviation of $\mathcal{A}_\perp(\phi)$ from the sinusoidal dependence is induced by the NB contribution.  If the spin vector $s_\perp$ is parallel to the normal vector $\mathbf{n}$, then only the $a_1$ component contributes to the spin asymmetry. In this case 
\begin{equation}
    \mathcal{A}_{\perp}(\phi=0^{\circ})  \sim  \mathcal{M}_{RES} \mathcal{M}_{NB}^*.
\end{equation}
The above property is illustrated in Fig.~\ref{Fig_asymmetry_Phi0chan0}, where we plot the decomposition of $\mathcal{A}_{\perp}(\phi=0^{\circ})$ into contributions from various interferences between diagrams. 

In Fig.~\ref{Fig_asymmetry_phi0=0} (top panel) we plot $\mathcal{A}_{\perp}(\phi=0^{\circ})$  as a function of energy. It is seen that the asymmetry is a small but nonvanishing function of the energy. The asymmetry takes the largest values when $s_\perp$ is perpendicular ($\phi = 90^{\circ}$) to the normal vector $\mathbf{n}$; see Fig.~\ref{Fig_asymmetry_phi0=0} (bottom panel).

\section{Summary}
\label{Sec_summary}
 
 Single pion production in the neutrino (antineutrino) scattering off the polarized target has been  discussed. Two polarizations of the target have been considered, namely, longitudinal and perpendicular to the neutrino beam. In both cases, the spin asymmetry has been calculated within two different SPP models. It is  demonstrated, in several examples, that  the spin asymmetry is sensitive to the nonresonant background contribution. Moreover, it is shown that when polarization of the target is parallel to the normal of the scattering plane, the asymmetry is given by the interference between the resonance and the nonresonance diagrams. 
 
 In summary, the spin asymmetry contains additional, with respect to spin averaged cross section measurements, information about the SPP dynamics, which  can be utilized to constrain significantly the single pion production models.   
 \bigbreak
 \bigbreak
 \bigbreak

 \section*{Acknowledgments}
The scattering amplitudes and cross sections have been calculated using symbolic programming language FORM~\cite{Vermaseren:2000nd}.
  \bigbreak
 	The calculations have been carried out in the Wroclaw Centre for Networking and Supercomputing \cite{wcss},
 	Grant No. 268.

\normalem
\bibliographystyle{apsrev4-1}
\bibliography{bibdrat,bibdratbook,bibmoje}

\begin{thebibliography}{54}%
\makeatletter
\providecommand \@ifxundefined [1]{%
 \@ifx{#1\undefined}
}%
\providecommand \@ifnum [1]{%
 \ifnum #1\expandafter \@firstoftwo
 \else \expandafter \@secondoftwo
 \fi
}%
\providecommand \@ifx [1]{%
 \ifx #1\expandafter \@firstoftwo
 \else \expandafter \@secondoftwo
 \fi
}%
\providecommand \natexlab [1]{#1}%
\providecommand \enquote  [1]{``#1''}%
\providecommand \bibnamefont  [1]{#1}%
\providecommand \bibfnamefont [1]{#1}%
\providecommand \citenamefont [1]{#1}%
\providecommand \href@noop [0]{\@secondoftwo}%
\providecommand \href [0]{\begingroup \@sanitize@url \@href}%
\providecommand \@href[1]{\@@startlink{#1}\@@href}%
\providecommand \@@href[1]{\endgroup#1\@@endlink}%
\providecommand \@sanitize@url [0]{\catcode `\\12\catcode `\$12\catcode
  `\&12\catcode `\#12\catcode `\^12\catcode `\_12\catcode `\%12\relax}%
\providecommand \@@startlink[1]{}%
\providecommand \@@endlink[0]{}%
\providecommand \url  [0]{\begingroup\@sanitize@url \@url }%
\providecommand \@url [1]{\endgroup\@href {#1}{\urlprefix }}%
\providecommand \urlprefix  [0]{URL }%
\providecommand \Eprint [0]{\href }%
\providecommand \doibase [0]{http://dx.doi.org/}%
\providecommand \selectlanguage [0]{\@gobble}%
\providecommand \bibinfo  [0]{\@secondoftwo}%
\providecommand \bibfield  [0]{\@secondoftwo}%
\providecommand \translation [1]{[#1]}%
\providecommand \BibitemOpen [0]{}%
\providecommand \bibitemStop [0]{}%
\providecommand \bibitemNoStop [0]{.\EOS\space}%
\providecommand \EOS [0]{\spacefactor3000\relax}%
\providecommand \BibitemShut  [1]{\csname bibitem#1\endcsname}%
\let\auto@bib@innerbib\@empty
\bibitem [{\citenamefont {Wascko}(2018)}]{wascko_morgan_2018_1286752}%
  \BibitemOpen
  \bibfield  {author} {\bibinfo {author} {\bibfnamefont {M.}~\bibnamefont
  {Wascko}},\ }\href {\doibase 10.5281/zenodo.1286752} {\enquote {\bibinfo
  {title} {T2k status, results, and plans},}\ } (\bibinfo {year}
  {2018})\BibitemShut {NoStop}%
\bibitem [{\citenamefont {Valle}(2018)}]{Valle:2018pgs}%
  \BibitemOpen
  \bibfield  {author} {\bibinfo {author} {\bibfnamefont {J.~W.~F.}\
  \bibnamefont {Valle}}\ }(\bibinfo {year} {2018})\ \Eprint
  {http://arxiv.org/abs/1812.07945} {arXiv:1812.07945 [hep-ph]} \BibitemShut
  {NoStop}%
\bibitem [{\citenamefont {Sanchez}(2018)}]{sanchez_mayly_2018_1286758}%
  \BibitemOpen
  \bibfield  {author} {\bibinfo {author} {\bibfnamefont {M.}~\bibnamefont
  {Sanchez}},\ }\href {\doibase 10.5281/zenodo.1286758} {\enquote {\bibinfo
  {title} {Nova results and prospects},}\ } (\bibinfo {year}
  {2018})\BibitemShut {NoStop}%
\bibitem [{\citenamefont
  {Mosel}(2016)}]{doi:10.1146/annurev-nucl-102115-044720}%
  \BibitemOpen
  \bibfield  {author} {\bibinfo {author} {\bibfnamefont {U.}~\bibnamefont
  {Mosel}},\ }\href {\doibase 10.1146/annurev-nucl-102115-044720} {\bibfield
  {journal} {\bibinfo  {journal} {{Annual Review of Nuclear and Particle
  Science}}\ }\textbf {\bibinfo {volume} {66}},\ \bibinfo {pages} {171}
  (\bibinfo {year} {2016})}\BibitemShut {NoStop}%
\bibitem [{\citenamefont {Sánchez}(2018)}]{sanchez_federico_2018_1295707}%
  \BibitemOpen
  \bibfield  {author} {\bibinfo {author} {\bibfnamefont {F.}~\bibnamefont
  {Sánchez}},\ }\href {\doibase 10.5281/zenodo.1295707} {\enquote {\bibinfo
  {title} {New neutrino cross-section measurements in t2k},}\ } (\bibinfo
  {year} {2018})\BibitemShut {NoStop}%
\bibitem [{\citenamefont {Adler}(1968)}]{Adler:1968tw}%
  \BibitemOpen
  \bibfield  {author} {\bibinfo {author} {\bibfnamefont {S.~L.}\ \bibnamefont
  {Adler}},\ }\href {\doibase 10.1016/0003-4916(68)90278-9} {\bibfield
  {journal} {\bibinfo  {journal} {Annals Phys.}\ }\textbf {\bibinfo {volume}
  {50}},\ \bibinfo {pages} {189} (\bibinfo {year} {1968})},\ \bibinfo {note}
  {[,225(1968)]}\BibitemShut {NoStop}%
\bibitem [{\citenamefont {Rein}\ and\ \citenamefont
  {Sehgal}(1981)}]{Rein:1980wg}%
  \BibitemOpen
  \bibfield  {author} {\bibinfo {author} {\bibfnamefont {D.}~\bibnamefont
  {Rein}}\ and\ \bibinfo {author} {\bibfnamefont {L.~M.}\ \bibnamefont
  {Sehgal}},\ }\href {\doibase 10.1016/0003-4916(81)90242-6} {\bibfield
  {journal} {\bibinfo  {journal} {Annals Phys.}\ }\textbf {\bibinfo {volume}
  {133}},\ \bibinfo {pages} {79} (\bibinfo {year} {1981})}\BibitemShut
  {NoStop}%
\bibitem [{\citenamefont {Fogli}\ and\ \citenamefont
  {Nardulli}(1979)}]{Fogli:1979cz}%
  \BibitemOpen
  \bibfield  {author} {\bibinfo {author} {\bibfnamefont {G.~L.}\ \bibnamefont
  {Fogli}}\ and\ \bibinfo {author} {\bibfnamefont {G.}~\bibnamefont
  {Nardulli}},\ }\href {\doibase 10.1016/0550-3213(79)90233-5} {\bibfield
  {journal} {\bibinfo  {journal} {Nucl. Phys.}\ }\textbf {\bibinfo {volume}
  {B160}},\ \bibinfo {pages} {116} (\bibinfo {year} {1979})}\BibitemShut
  {NoStop}%
\bibitem [{\citenamefont {Gershtein}\ \emph {et~al.}(1980)\citenamefont
  {Gershtein}, \citenamefont {Komachenko},\ and\ \citenamefont
  {Khlopov}}]{Gershtein:1980vd}%
  \BibitemOpen
  \bibfield  {author} {\bibinfo {author} {\bibfnamefont {S.~S.}\ \bibnamefont
  {Gershtein}}, \bibinfo {author} {\bibfnamefont {{\relax Yu}.~{\relax Ya}.}\
  \bibnamefont {Komachenko}}, \ and\ \bibinfo {author} {\bibfnamefont
  {M.~{\relax Yu}.}\ \bibnamefont {Khlopov}},\ }\href@noop {} {\bibfield
  {journal} {\bibinfo  {journal} {Sov. J. Nucl. Phys.}\ }\textbf {\bibinfo
  {volume} {32}},\ \bibinfo {pages} {861} (\bibinfo {year} {1980})},\ \bibinfo
  {note} {[Yad. Fiz.32(1980)]}\BibitemShut {NoStop}%
\bibitem [{\citenamefont {Rein}(1987)}]{Rein:1987cb}%
  \BibitemOpen
  \bibfield  {author} {\bibinfo {author} {\bibfnamefont {D.}~\bibnamefont
  {Rein}},\ }\href {\doibase 10.1007/BF01561054} {\bibfield  {journal}
  {\bibinfo  {journal} {Z. Phys.}\ }\textbf {\bibinfo {volume} {C35}},\
  \bibinfo {pages} {43} (\bibinfo {year} {1987})}\BibitemShut {NoStop}%
\bibitem [{\citenamefont {Sato}\ \emph {et~al.}(2003)\citenamefont {Sato},
  \citenamefont {Uno},\ and\ \citenamefont {Lee}}]{Sato:2003rq}%
  \BibitemOpen
  \bibfield  {author} {\bibinfo {author} {\bibfnamefont {T.}~\bibnamefont
  {Sato}}, \bibinfo {author} {\bibfnamefont {D.}~\bibnamefont {Uno}}, \ and\
  \bibinfo {author} {\bibfnamefont {T.~S.~H.}\ \bibnamefont {Lee}},\ }\href
  {\doibase 10.1103/PhysRevC.67.065201} {\bibfield  {journal} {\bibinfo
  {journal} {Phys. Rev.}\ }\textbf {\bibinfo {volume} {C67}},\ \bibinfo {pages}
  {065201} (\bibinfo {year} {2003})},\ \Eprint
  {http://arxiv.org/abs/nucl-th/0303050} {arXiv:nucl-th/0303050 [nucl-th]}
  \BibitemShut {NoStop}%
\bibitem [{\citenamefont {Hernandez}\ \emph
  {et~al.}(2007{\natexlab{a}})\citenamefont {Hernandez}, \citenamefont
  {Nieves},\ and\ \citenamefont {Valverde}}]{Hernandez:2006yg}%
  \BibitemOpen
  \bibfield  {author} {\bibinfo {author} {\bibfnamefont {E.}~\bibnamefont
  {Hernandez}}, \bibinfo {author} {\bibfnamefont {J.}~\bibnamefont {Nieves}}, \
  and\ \bibinfo {author} {\bibfnamefont {M.}~\bibnamefont {Valverde}},\ }\href
  {\doibase 10.1016/j.physletb.2007.02.051} {\bibfield  {journal} {\bibinfo
  {journal} {Phys. Lett.}\ }\textbf {\bibinfo {volume} {B647}},\ \bibinfo
  {pages} {452} (\bibinfo {year} {2007}{\natexlab{a}})},\ \Eprint
  {http://arxiv.org/abs/hep-ph/0608119} {arXiv:hep-ph/0608119 [hep-ph]}
  \BibitemShut {NoStop}%
\bibitem [{\citenamefont {Graczyk}\ and\ \citenamefont
  {Sobczyk}(2008)}]{Graczyk:2007bc}%
  \BibitemOpen
  \bibfield  {author} {\bibinfo {author} {\bibfnamefont {K.~M.}\ \bibnamefont
  {Graczyk}}\ and\ \bibinfo {author} {\bibfnamefont {J.~T.}\ \bibnamefont
  {Sobczyk}},\ }\href {\doibase 10.1103/PhysRevD.79.079903,
  10.1103/PhysRevD.77.053001} {\bibfield  {journal} {\bibinfo  {journal} {Phys.
  Rev.}\ }\textbf {\bibinfo {volume} {D77}},\ \bibinfo {pages} {053001}
  (\bibinfo {year} {2008})},\ \bibinfo {note} {[Erratum: Phys.
  Rev.D79,079903(2009)]},\ \Eprint {http://arxiv.org/abs/0707.3561}
  {arXiv:0707.3561 [hep-ph]} \BibitemShut {NoStop}%
\bibitem [{\citenamefont {Leitner}\ \emph {et~al.}(2009)\citenamefont
  {Leitner}, \citenamefont {Buss}, \citenamefont {Alvarez-Ruso},\ and\
  \citenamefont {Mosel}}]{Leitner:2008ue}%
  \BibitemOpen
  \bibfield  {author} {\bibinfo {author} {\bibfnamefont {T.}~\bibnamefont
  {Leitner}}, \bibinfo {author} {\bibfnamefont {O.}~\bibnamefont {Buss}},
  \bibinfo {author} {\bibfnamefont {L.}~\bibnamefont {Alvarez-Ruso}}, \ and\
  \bibinfo {author} {\bibfnamefont {U.}~\bibnamefont {Mosel}},\ }\href
  {\doibase 10.1103/PhysRevC.79.034601} {\bibfield  {journal} {\bibinfo
  {journal} {Phys. Rev.}\ }\textbf {\bibinfo {volume} {C79}},\ \bibinfo {pages}
  {034601} (\bibinfo {year} {2009})},\ \Eprint {http://arxiv.org/abs/0812.0587}
  {arXiv:0812.0587 [nucl-th]} \BibitemShut {NoStop}%
\bibitem [{\citenamefont {Graczyk}\ \emph {et~al.}(2009)\citenamefont
  {Graczyk}, \citenamefont {Kielczewska}, \citenamefont {Przewlocki},\ and\
  \citenamefont {Sobczyk}}]{Graczyk:2009qm}%
  \BibitemOpen
  \bibfield  {author} {\bibinfo {author} {\bibfnamefont {K.~M.}\ \bibnamefont
  {Graczyk}}, \bibinfo {author} {\bibfnamefont {D.}~\bibnamefont
  {Kielczewska}}, \bibinfo {author} {\bibfnamefont {P.}~\bibnamefont
  {Przewlocki}}, \ and\ \bibinfo {author} {\bibfnamefont {J.~T.}\ \bibnamefont
  {Sobczyk}},\ }\href {\doibase 10.1103/PhysRevD.80.093001} {\bibfield
  {journal} {\bibinfo  {journal} {Phys. Rev.}\ }\textbf {\bibinfo {volume}
  {D80}},\ \bibinfo {pages} {093001} (\bibinfo {year} {2009})},\ \Eprint
  {http://arxiv.org/abs/0908.2175} {arXiv:0908.2175 [hep-ph]} \BibitemShut
  {NoStop}%
\bibitem [{\citenamefont {Nakamura}\ \emph {et~al.}(2015)\citenamefont
  {Nakamura}, \citenamefont {Kamano},\ and\ \citenamefont
  {Sato}}]{Nakamura:2015rta}%
  \BibitemOpen
  \bibfield  {author} {\bibinfo {author} {\bibfnamefont {S.~X.}\ \bibnamefont
  {Nakamura}}, \bibinfo {author} {\bibfnamefont {H.}~\bibnamefont {Kamano}}, \
  and\ \bibinfo {author} {\bibfnamefont {T.}~\bibnamefont {Sato}},\ }\href
  {\doibase 10.1103/PhysRevD.92.074024} {\bibfield  {journal} {\bibinfo
  {journal} {Phys. Rev.}\ }\textbf {\bibinfo {volume} {D92}},\ \bibinfo {pages}
  {074024} (\bibinfo {year} {2015})},\ \Eprint
  {http://arxiv.org/abs/1506.03403} {arXiv:1506.03403 [hep-ph]} \BibitemShut
  {NoStop}%
\bibitem [{\citenamefont {Serot}\ and\ \citenamefont
  {Zhang}(2012)}]{Serot:2012rd}%
  \BibitemOpen
  \bibfield  {author} {\bibinfo {author} {\bibfnamefont {B.~D.}\ \bibnamefont
  {Serot}}\ and\ \bibinfo {author} {\bibfnamefont {X.}~\bibnamefont {Zhang}},\
  }\href {\doibase 10.1103/PhysRevC.86.015501} {\bibfield  {journal} {\bibinfo
  {journal} {Phys. Rev.}\ }\textbf {\bibinfo {volume} {C86}},\ \bibinfo {pages}
  {015501} (\bibinfo {year} {2012})},\ \Eprint {http://arxiv.org/abs/1206.3812}
  {arXiv:1206.3812 [nucl-th]} \BibitemShut {NoStop}%
\bibitem [{\citenamefont {Lalakulich}\ \emph {et~al.}(2010)\citenamefont
  {Lalakulich}, \citenamefont {Leitner}, \citenamefont {Buss},\ and\
  \citenamefont {Mosel}}]{Lalakulich:2010ss}%
  \BibitemOpen
  \bibfield  {author} {\bibinfo {author} {\bibfnamefont {O.}~\bibnamefont
  {Lalakulich}}, \bibinfo {author} {\bibfnamefont {T.}~\bibnamefont {Leitner}},
  \bibinfo {author} {\bibfnamefont {O.}~\bibnamefont {Buss}}, \ and\ \bibinfo
  {author} {\bibfnamefont {U.}~\bibnamefont {Mosel}},\ }\href {\doibase
  10.1103/PhysRevD.82.093001} {\bibfield  {journal} {\bibinfo  {journal} {Phys.
  Rev.}\ }\textbf {\bibinfo {volume} {D82}},\ \bibinfo {pages} {093001}
  (\bibinfo {year} {2010})},\ \Eprint {http://arxiv.org/abs/1007.0925}
  {arXiv:1007.0925 [hep-ph]} \BibitemShut {NoStop}%
\bibitem [{\citenamefont {Rafi~Alam}\ \emph {et~al.}(2016)\citenamefont
  {Rafi~Alam}, \citenamefont {Sajjad~Athar}, \citenamefont {Chauhan},\ and\
  \citenamefont {Singh}}]{Alam:2015gaa}%
  \BibitemOpen
  \bibfield  {author} {\bibinfo {author} {\bibfnamefont {M.}~\bibnamefont
  {Rafi~Alam}}, \bibinfo {author} {\bibfnamefont {M.}~\bibnamefont
  {Sajjad~Athar}}, \bibinfo {author} {\bibfnamefont {S.}~\bibnamefont
  {Chauhan}}, \ and\ \bibinfo {author} {\bibfnamefont {S.~K.}\ \bibnamefont
  {Singh}},\ }\href {\doibase 10.1142/S0218301316500105} {\bibfield  {journal}
  {\bibinfo  {journal} {Int. J. Mod. Phys.}\ }\textbf {\bibinfo {volume}
  {E25}},\ \bibinfo {pages} {1650010} (\bibinfo {year} {2016})},\ \Eprint
  {http://arxiv.org/abs/1509.08622} {arXiv:1509.08622 [hep-ph]} \BibitemShut
  {NoStop}%
\bibitem [{\citenamefont {Graczyk}\ \emph {et~al.}(2014)\citenamefont
  {Graczyk}, \citenamefont {Zmuda},\ and\ \citenamefont
  {Sobczyk}}]{Graczyk:2014dpa}%
  \BibitemOpen
  \bibfield  {author} {\bibinfo {author} {\bibfnamefont {K.~M.}\ \bibnamefont
  {Graczyk}}, \bibinfo {author} {\bibfnamefont {J.}~\bibnamefont {Zmuda}}, \
  and\ \bibinfo {author} {\bibfnamefont {J.~T.}\ \bibnamefont {Sobczyk}},\
  }\href {\doibase 10.1103/PhysRevD.90.093001} {\bibfield  {journal} {\bibinfo
  {journal} {Phys. Rev.}\ }\textbf {\bibinfo {volume} {D90}},\ \bibinfo {pages}
  {093001} (\bibinfo {year} {2014})},\ \Eprint {http://arxiv.org/abs/1407.5445}
  {arXiv:1407.5445 [hep-ph]} \BibitemShut {NoStop}%
\bibitem [{\citenamefont {Barbero}\ \emph {et~al.}(2008)\citenamefont
  {Barbero}, \citenamefont {Lopez~Castro},\ and\ \citenamefont
  {Mariano}}]{Barbero:2008zza}%
  \BibitemOpen
  \bibfield  {author} {\bibinfo {author} {\bibfnamefont {C.}~\bibnamefont
  {Barbero}}, \bibinfo {author} {\bibfnamefont {G.}~\bibnamefont
  {Lopez~Castro}}, \ and\ \bibinfo {author} {\bibfnamefont {A.}~\bibnamefont
  {Mariano}},\ }\href {\doibase 10.1016/j.physletb.2008.05.011} {\bibfield
  {journal} {\bibinfo  {journal} {Phys. Lett.}\ }\textbf {\bibinfo {volume}
  {B664}},\ \bibinfo {pages} {70} (\bibinfo {year} {2008})}\BibitemShut
  {NoStop}%
\bibitem [{\citenamefont {Alvarez-Ruso}\ \emph {et~al.}(2016)\citenamefont
  {Alvarez-Ruso}, \citenamefont {Hernandez}, \citenamefont {Nieves},\ and\
  \citenamefont {Vicente~Vacas}}]{Alvarez-Ruso:2015eva}%
  \BibitemOpen
  \bibfield  {author} {\bibinfo {author} {\bibfnamefont {L.}~\bibnamefont
  {Alvarez-Ruso}}, \bibinfo {author} {\bibfnamefont {E.}~\bibnamefont
  {Hernandez}}, \bibinfo {author} {\bibfnamefont {J.}~\bibnamefont {Nieves}}, \
  and\ \bibinfo {author} {\bibfnamefont {M.~J.}\ \bibnamefont
  {Vicente~Vacas}},\ }\href {\doibase 10.1103/PhysRevD.93.014016} {\bibfield
  {journal} {\bibinfo  {journal} {Phys. Rev.}\ }\textbf {\bibinfo {volume}
  {D93}},\ \bibinfo {pages} {014016} (\bibinfo {year} {2016})},\ \Eprint
  {http://arxiv.org/abs/1510.06266} {arXiv:1510.06266 [hep-ph]} \BibitemShut
  {NoStop}%
\bibitem [{\citenamefont {Gonzalez-Jimenez}\ \emph {et~al.}(2017)\citenamefont
  {Gonzalez-Jimenez}, \citenamefont {Jachowicz}, \citenamefont {Niewczas},
  \citenamefont {Nys}, \citenamefont {Pandey}, \citenamefont {Van~Cuyck},\ and\
  \citenamefont {Van~Dessel}}]{Gonzalez-Jimenez:2016qqq}%
  \BibitemOpen
  \bibfield  {author} {\bibinfo {author} {\bibfnamefont {R.}~\bibnamefont
  {Gonzalez-Jimenez}}, \bibinfo {author} {\bibfnamefont {N.}~\bibnamefont
  {Jachowicz}}, \bibinfo {author} {\bibfnamefont {K.}~\bibnamefont {Niewczas}},
  \bibinfo {author} {\bibfnamefont {J.}~\bibnamefont {Nys}}, \bibinfo {author}
  {\bibfnamefont {V.}~\bibnamefont {Pandey}}, \bibinfo {author} {\bibfnamefont
  {T.}~\bibnamefont {Van~Cuyck}}, \ and\ \bibinfo {author} {\bibfnamefont
  {N.}~\bibnamefont {Van~Dessel}},\ }\href {\doibase
  10.1103/PhysRevD.95.113007} {\bibfield  {journal} {\bibinfo  {journal} {Phys.
  Rev.}\ }\textbf {\bibinfo {volume} {D95}},\ \bibinfo {pages} {113007}
  (\bibinfo {year} {2017})},\ \Eprint {http://arxiv.org/abs/1612.05511}
  {arXiv:1612.05511 [nucl-th]} \BibitemShut {NoStop}%
\bibitem [{\citenamefont {Hernandez}\ and\ \citenamefont
  {Nieves}(2017)}]{Hernandez:2016yfb}%
  \BibitemOpen
  \bibfield  {author} {\bibinfo {author} {\bibfnamefont {E.}~\bibnamefont
  {Hernandez}}\ and\ \bibinfo {author} {\bibfnamefont {J.}~\bibnamefont
  {Nieves}},\ }\href {\doibase 10.1103/PhysRevD.95.053007} {\bibfield
  {journal} {\bibinfo  {journal} {Phys. Rev.}\ }\textbf {\bibinfo {volume}
  {D95}},\ \bibinfo {pages} {053007} (\bibinfo {year} {2017})},\ \Eprint
  {http://arxiv.org/abs/1612.02343} {arXiv:1612.02343 [hep-ph]} \BibitemShut
  {NoStop}%
\bibitem [{\citenamefont {Yao}\ \emph {et~al.}(2018)\citenamefont {Yao},
  \citenamefont {Alvarez-Ruso}, \citenamefont {Blin},\ and\ \citenamefont
  {Vicente~Vacas}}]{Yao:2018pzc}%
  \BibitemOpen
  \bibfield  {author} {\bibinfo {author} {\bibfnamefont {D.-L.}\ \bibnamefont
  {Yao}}, \bibinfo {author} {\bibfnamefont {L.}~\bibnamefont {Alvarez-Ruso}},
  \bibinfo {author} {\bibfnamefont {A.~N.~H.}\ \bibnamefont {Blin}}, \ and\
  \bibinfo {author} {\bibfnamefont {M.~J.}\ \bibnamefont {Vicente~Vacas}},\
  }\href@noop {} {\  (\bibinfo {year} {2018})},\ \Eprint
  {http://arxiv.org/abs/1806.09364} {arXiv:1806.09364 [hep-ph]} \BibitemShut
  {NoStop}%
\bibitem [{\citenamefont {Yao}\ \emph {et~al.}(2019)\citenamefont {Yao},
  \citenamefont {Alvarez-Ruso},\ and\ \citenamefont
  {Vicente~Vacas}}]{Yao:2019avf}%
  \BibitemOpen
  \bibfield  {author} {\bibinfo {author} {\bibfnamefont {D.-L.}\ \bibnamefont
  {Yao}}, \bibinfo {author} {\bibfnamefont {L.}~\bibnamefont {Alvarez-Ruso}}, \
  and\ \bibinfo {author} {\bibfnamefont {M.~J.}\ \bibnamefont
  {Vicente~Vacas}},\ }\href@noop {} {\  (\bibinfo {year} {2019})},\ \Eprint
  {http://arxiv.org/abs/1901.00773} {arXiv:1901.00773 [hep-ph]} \BibitemShut
  {NoStop}%
\bibitem [{\citenamefont {Radecky}\ \emph {et~al.}(1982)\citenamefont
  {Radecky}, \citenamefont {Barnes}, \citenamefont {Carmony}, \citenamefont
  {Garfinkel}, \citenamefont {Derrick}, \citenamefont {Fernandez},
  \citenamefont {Hyman},\ and\ \citenamefont {Levman~{\it et
  al.}}}]{Radecky:1981fn}%
  \BibitemOpen
  \bibfield  {author} {\bibinfo {author} {\bibfnamefont {G.~M.}\ \bibnamefont
  {Radecky}}, \bibinfo {author} {\bibfnamefont {V.~E.}\ \bibnamefont {Barnes}},
  \bibinfo {author} {\bibfnamefont {D.~D.}\ \bibnamefont {Carmony}}, \bibinfo
  {author} {\bibfnamefont {A.~F.}\ \bibnamefont {Garfinkel}}, \bibinfo {author}
  {\bibfnamefont {M.}~\bibnamefont {Derrick}}, \bibinfo {author} {\bibfnamefont
  {E.}~\bibnamefont {Fernandez}}, \bibinfo {author} {\bibfnamefont
  {L.}~\bibnamefont {Hyman}}, \ and\ \bibinfo {author} {\bibfnamefont
  {G.}~\bibnamefont {Levman~{\it et al.}}},\ }\href@noop {} {\bibfield
  {journal} {\bibinfo  {journal} {Phys.\ Rev.\ D}\ }\textbf {\bibinfo {volume}
  {{\bf 26}}},\ \bibinfo {pages} {3297} (\bibinfo {year} {1982})},\ \bibinfo
  {note} {[Erratum-ibid.\ D {\bf 26} (1982) 3297]}\BibitemShut {NoStop}%
\bibitem [{\citenamefont {Kitagaki}\ \emph {et~al.}(1986)\citenamefont
  {Kitagaki}, \citenamefont {Yuta}, \citenamefont {Tanaka}, \citenamefont
  {Yamaguchi}, \citenamefont {Abe} \emph {et~al.}}]{Kitagaki:1986ct}%
  \BibitemOpen
  \bibfield  {author} {\bibinfo {author} {\bibfnamefont {T.}~\bibnamefont
  {Kitagaki}}, \bibinfo {author} {\bibfnamefont {H.}~\bibnamefont {Yuta}},
  \bibinfo {author} {\bibfnamefont {S.}~\bibnamefont {Tanaka}}, \bibinfo
  {author} {\bibfnamefont {A.}~\bibnamefont {Yamaguchi}}, \bibinfo {author}
  {\bibfnamefont {K.}~\bibnamefont {Abe}},  \emph {et~al.},\ }\href {\doibase
  10.1103/PhysRevD.34.2554} {\bibfield  {journal} {\bibinfo  {journal} {Phys.
  Rev. D}\ }\textbf {\bibinfo {volume} {34}},\ \bibinfo {pages} {2554}
  (\bibinfo {year} {1986})}\BibitemShut {NoStop}%
\bibitem [{\citenamefont {Rodriguez}\ \emph {et~al.}(2008)\citenamefont
  {Rodriguez} \emph {et~al.}}]{Rodriguez:2008aa}%
  \BibitemOpen
  \bibfield  {author} {\bibinfo {author} {\bibfnamefont {A.}~\bibnamefont
  {Rodriguez}} \emph {et~al.} (\bibinfo {collaboration} {K2K}),\ }\href
  {\doibase 10.1103/PhysRevD.78.032003} {\bibfield  {journal} {\bibinfo
  {journal} {Phys. Rev.}\ }\textbf {\bibinfo {volume} {D78}},\ \bibinfo {pages}
  {032003} (\bibinfo {year} {2008})},\ \Eprint {http://arxiv.org/abs/0805.0186}
  {arXiv:0805.0186 [hep-ex]} \BibitemShut {NoStop}%
\bibitem [{\citenamefont {Aguilar-Arevalo}\ \emph {et~al.}(2011)\citenamefont
  {Aguilar-Arevalo} \emph {et~al.}}]{AguilarArevalo:2010bm}%
  \BibitemOpen
  \bibfield  {author} {\bibinfo {author} {\bibfnamefont {A.~A.}\ \bibnamefont
  {Aguilar-Arevalo}} \emph {et~al.} (\bibinfo {collaboration} {MiniBooNE}),\
  }\href {\doibase 10.1103/PhysRevD.83.052007} {\bibfield  {journal} {\bibinfo
  {journal} {Phys. Rev.}\ }\textbf {\bibinfo {volume} {D83}},\ \bibinfo {pages}
  {052007} (\bibinfo {year} {2011})},\ \Eprint {http://arxiv.org/abs/1011.3572}
  {arXiv:1011.3572 [hep-ex]} \BibitemShut {NoStop}%
\bibitem [{\citenamefont {McGivern}\ \emph {et~al.}(2016)\citenamefont
  {McGivern} \emph {et~al.}}]{McGivern:2016bwh}%
  \BibitemOpen
  \bibfield  {author} {\bibinfo {author} {\bibfnamefont {C.~L.}\ \bibnamefont
  {McGivern}} \emph {et~al.} (\bibinfo {collaboration} {MINERvA}),\ }\href
  {\doibase 10.1103/PhysRevD.94.052005} {\bibfield  {journal} {\bibinfo
  {journal} {Phys. Rev.}\ }\textbf {\bibinfo {volume} {D94}},\ \bibinfo {pages}
  {052005} (\bibinfo {year} {2016})},\ \Eprint
  {http://arxiv.org/abs/1606.07127} {arXiv:1606.07127 [hep-ex]} \BibitemShut
  {NoStop}%
\bibitem [{\citenamefont {Abe}\ \emph {et~al.}(2017)\citenamefont {Abe} \emph
  {et~al.}}]{Abe:2016aoo}%
  \BibitemOpen
  \bibfield  {author} {\bibinfo {author} {\bibfnamefont {K.}~\bibnamefont
  {Abe}} \emph {et~al.} (\bibinfo {collaboration} {T2K}),\ }\href {\doibase
  10.1103/PhysRevD.95.012010} {\bibfield  {journal} {\bibinfo  {journal} {Phys.
  Rev.}\ }\textbf {\bibinfo {volume} {D95}},\ \bibinfo {pages} {012010}
  (\bibinfo {year} {2017})},\ \Eprint {http://arxiv.org/abs/1605.07964}
  {arXiv:1605.07964 [hep-ex]} \BibitemShut {NoStop}%
\bibitem [{\citenamefont {Graczyk}\ and\ \citenamefont
  {Kowal}(2018)}]{Graczyk:2017rti}%
  \BibitemOpen
  \bibfield  {author} {\bibinfo {author} {\bibfnamefont {K.~M.}\ \bibnamefont
  {Graczyk}}\ and\ \bibinfo {author} {\bibfnamefont {B.~E.}\ \bibnamefont
  {Kowal}},\ }\href {\doibase 10.1103/PhysRevD.97.013001} {\bibfield  {journal}
  {\bibinfo  {journal} {Phys. Rev.}\ }\textbf {\bibinfo {volume} {D97}},\
  \bibinfo {pages} {013001} (\bibinfo {year} {2018})},\ \Eprint
  {http://arxiv.org/abs/1711.04868} {arXiv:1711.04868 [hep-ph]} \BibitemShut
  {NoStop}%
\bibitem [{\citenamefont {Adler}(1963)}]{Adler1963}%
  \BibitemOpen
  \bibfield  {author} {\bibinfo {author} {\bibfnamefont {S.~L.}\ \bibnamefont
  {Adler}},\ }\href {\doibase 10.1007/BF02828811} {\bibfield  {journal}
  {\bibinfo  {journal} {Il Nuovo Cimento (1955-1965)}\ }\textbf {\bibinfo
  {volume} {30}},\ \bibinfo {pages} {1020} (\bibinfo {year}
  {1963})}\BibitemShut {NoStop}%
\bibitem [{\citenamefont {Pais}(1971)}]{Pais:1971er}%
  \BibitemOpen
  \bibfield  {author} {\bibinfo {author} {\bibfnamefont {A.}~\bibnamefont
  {Pais}},\ }\href {\doibase 10.1016/0003-4916(71)90018-2} {\bibfield
  {journal} {\bibinfo  {journal} {Annals Phys.}\ }\textbf {\bibinfo {volume}
  {63}},\ \bibinfo {pages} {361} (\bibinfo {year} {1971})}\BibitemShut
  {NoStop}%
\bibitem [{\citenamefont {Llewellyn~Smith}(1972)}]{LlewellynSmith:1971uhs}%
  \BibitemOpen
  \bibfield  {author} {\bibinfo {author} {\bibfnamefont {C.~H.}\ \bibnamefont
  {Llewellyn~Smith}},\ }\bibfield  {booktitle} {\emph {\bibinfo {booktitle}
  {{Gauge Theories and Neutrino Physics, Jacob, 1978:0175}}},\ }\href {\doibase
  10.1016/0370-1573(72)90010-5} {\bibfield  {journal} {\bibinfo  {journal}
  {Phys. Rept.}\ }\textbf {\bibinfo {volume} {3}},\ \bibinfo {pages} {261}
  (\bibinfo {year} {1972})}\BibitemShut {NoStop}%
\bibitem [{\citenamefont {Kuzmin}\ \emph {et~al.}(2004)\citenamefont {Kuzmin},
  \citenamefont {Lyubushkin},\ and\ \citenamefont {Naumov}}]{Kuzmin:2003ji}%
  \BibitemOpen
  \bibfield  {author} {\bibinfo {author} {\bibfnamefont {K.~S.}\ \bibnamefont
  {Kuzmin}}, \bibinfo {author} {\bibfnamefont {V.~V.}\ \bibnamefont
  {Lyubushkin}}, \ and\ \bibinfo {author} {\bibfnamefont {V.~A.}\ \bibnamefont
  {Naumov}},\ }\bibfield  {booktitle} {\emph {\bibinfo {booktitle}
  {{Proceedings, 10th Advanced Research Workshop on High-Energy Spin Physics
  (SPIN-03): Dubna, Russia, September 16-20, 2003}}},\ }\href {\doibase
  10.1142/S0217732304016172} {\bibfield  {journal} {\bibinfo  {journal} {Mod.
  Phys. Lett.}\ }\textbf {\bibinfo {volume} {A19}},\ \bibinfo {pages} {2815}
  (\bibinfo {year} {2004})},\ \bibinfo {note} {[,125(2003)]},\ \Eprint
  {http://arxiv.org/abs/hep-ph/0312107} {arXiv:hep-ph/0312107 [hep-ph]}
  \BibitemShut {NoStop}%
\bibitem [{\citenamefont {Hagiwara}\ \emph {et~al.}(2003)\citenamefont
  {Hagiwara}, \citenamefont {Mawatari},\ and\ \citenamefont
  {Yokoya}}]{Hagiwara:2003di}%
  \BibitemOpen
  \bibfield  {author} {\bibinfo {author} {\bibfnamefont {K.}~\bibnamefont
  {Hagiwara}}, \bibinfo {author} {\bibfnamefont {K.}~\bibnamefont {Mawatari}},
  \ and\ \bibinfo {author} {\bibfnamefont {H.}~\bibnamefont {Yokoya}},\ }\href
  {\doibase 10.1016/S0550-3213(03)00575-3} {\bibfield  {journal} {\bibinfo
  {journal} {Nucl. Phys.}\ }\textbf {\bibinfo {volume} {B668}},\ \bibinfo
  {pages} {364} (\bibinfo {year} {2003})},\ \bibinfo {note} {[Erratum: Nucl.
  Phys.B701,405(2004)]},\ \Eprint {http://arxiv.org/abs/hep-ph/0305324}
  {arXiv:hep-ph/0305324 [hep-ph]} \BibitemShut {NoStop}%
\bibitem [{\citenamefont {Graczyk}(2005)}]{Graczyk:2004uy}%
  \BibitemOpen
  \bibfield  {author} {\bibinfo {author} {\bibfnamefont {K.~M.}\ \bibnamefont
  {Graczyk}},\ }\href {\doibase 10.1016/j.nuclphysa.2004.10.029} {\bibfield
  {journal} {\bibinfo  {journal} {Nucl. Phys.}\ }\textbf {\bibinfo {volume}
  {A748}},\ \bibinfo {pages} {313} (\bibinfo {year} {2005})},\ \Eprint
  {http://arxiv.org/abs/hep-ph/0407275} {arXiv:hep-ph/0407275 [hep-ph]}
  \BibitemShut {NoStop}%
\bibitem [{\citenamefont {Kuzmin}\ \emph {et~al.}(2005)\citenamefont {Kuzmin},
  \citenamefont {Lyubushkin},\ and\ \citenamefont {Naumov}}]{Kuzmin:2004yb}%
  \BibitemOpen
  \bibfield  {author} {\bibinfo {author} {\bibfnamefont {K.~S.}\ \bibnamefont
  {Kuzmin}}, \bibinfo {author} {\bibfnamefont {V.~V.}\ \bibnamefont
  {Lyubushkin}}, \ and\ \bibinfo {author} {\bibfnamefont {V.~A.}\ \bibnamefont
  {Naumov}},\ }\bibfield  {booktitle} {\emph {\bibinfo {booktitle}
  {{Proceedings, 3rd International Workshop on Neutrino-nucleus interactions in
  the few GeV region (NUINT 04): Assergi, Italy, March 17-21, 2004}}},\ }\href
  {\doibase 10.1016/j.nuclphysbps.2004.11.221} {\bibfield  {journal} {\bibinfo
  {journal} {Nucl. Phys. Proc. Suppl.}\ }\textbf {\bibinfo {volume} {139}},\
  \bibinfo {pages} {154} (\bibinfo {year} {2005})},\ \bibinfo {note}
  {[,154(2004)]},\ \Eprint {http://arxiv.org/abs/hep-ph/0408107}
  {arXiv:hep-ph/0408107 [hep-ph]} \BibitemShut {NoStop}%
\bibitem [{\citenamefont {Bilenky}\ and\ \citenamefont
  {Christova}(2013{\natexlab{a}})}]{Bilenky:2013iua}%
  \BibitemOpen
  \bibfield  {author} {\bibinfo {author} {\bibfnamefont {S.~M.}\ \bibnamefont
  {Bilenky}}\ and\ \bibinfo {author} {\bibfnamefont {E.}~\bibnamefont
  {Christova}},\ }\href {\doibase 10.1134/S154747711307011X} {\bibfield
  {journal} {\bibinfo  {journal} {Phys. Part. Nucl. Lett.}\ }\textbf {\bibinfo
  {volume} {10}},\ \bibinfo {pages} {651} (\bibinfo {year}
  {2013}{\natexlab{a}})},\ \Eprint {http://arxiv.org/abs/1307.7275}
  {arXiv:1307.7275 [hep-ph]} \BibitemShut {NoStop}%
\bibitem [{\citenamefont {Bilenky}\ and\ \citenamefont
  {Christova}(2013{\natexlab{b}})}]{Bilenky:2013fra}%
  \BibitemOpen
  \bibfield  {author} {\bibinfo {author} {\bibfnamefont {S.~M.}\ \bibnamefont
  {Bilenky}}\ and\ \bibinfo {author} {\bibfnamefont {E.}~\bibnamefont
  {Christova}},\ }\href {\doibase 10.1088/0954-3899/40/7/075004} {\bibfield
  {journal} {\bibinfo  {journal} {J. Phys.}\ }\textbf {\bibinfo {volume}
  {G40}},\ \bibinfo {pages} {075004} (\bibinfo {year} {2013}{\natexlab{b}})},\
  \Eprint {http://arxiv.org/abs/1303.3710} {arXiv:1303.3710 [hep-ph]}
  \BibitemShut {NoStop}%
\bibitem [{\citenamefont {Akbar}\ \emph {et~al.}(2016)\citenamefont {Akbar},
  \citenamefont {Rafi~Alam}, \citenamefont {Sajjad~Athar},\ and\ \citenamefont
  {Singh}}]{Akbar:2016awk}%
  \BibitemOpen
  \bibfield  {author} {\bibinfo {author} {\bibfnamefont {F.}~\bibnamefont
  {Akbar}}, \bibinfo {author} {\bibfnamefont {M.}~\bibnamefont {Rafi~Alam}},
  \bibinfo {author} {\bibfnamefont {M.}~\bibnamefont {Sajjad~Athar}}, \ and\
  \bibinfo {author} {\bibfnamefont {S.~K.}\ \bibnamefont {Singh}},\ }\href
  {\doibase 10.1103/PhysRevD.94.114031} {\bibfield  {journal} {\bibinfo
  {journal} {Phys. Rev.}\ }\textbf {\bibinfo {volume} {D94}},\ \bibinfo {pages}
  {114031} (\bibinfo {year} {2016})},\ \Eprint
  {http://arxiv.org/abs/1608.02103} {arXiv:1608.02103 [hep-ph]} \BibitemShut
  {NoStop}%
\bibitem [{\citenamefont {Akbar}\ \emph {et~al.}(2017)\citenamefont {Akbar},
  \citenamefont {Sajjad~Athar}, \citenamefont {Fatima},\ and\ \citenamefont
  {Singh}}]{Akbar:2017qsf}%
  \BibitemOpen
  \bibfield  {author} {\bibinfo {author} {\bibfnamefont {F.}~\bibnamefont
  {Akbar}}, \bibinfo {author} {\bibfnamefont {M.}~\bibnamefont {Sajjad~Athar}},
  \bibinfo {author} {\bibfnamefont {A.}~\bibnamefont {Fatima}}, \ and\ \bibinfo
  {author} {\bibfnamefont {S.~K.}\ \bibnamefont {Singh}},\ }\href {\doibase
  10.1140/epja/i2017-12340-4} {\bibfield  {journal} {\bibinfo  {journal} {Eur.
  Phys. J.}\ }\textbf {\bibinfo {volume} {A53}},\ \bibinfo {pages} {154}
  (\bibinfo {year} {2017})},\ \Eprint {http://arxiv.org/abs/1704.04580}
  {arXiv:1704.04580 [hep-ph]} \BibitemShut {NoStop}%
\bibitem [{\citenamefont {Fatima}\ \emph
  {et~al.}(2018{\natexlab{a}})\citenamefont {Fatima}, \citenamefont
  {Sajjad~Athar},\ and\ \citenamefont {Singh}}]{Fatima:2018gjy}%
  \BibitemOpen
  \bibfield  {author} {\bibinfo {author} {\bibfnamefont {A.}~\bibnamefont
  {Fatima}}, \bibinfo {author} {\bibfnamefont {M.}~\bibnamefont
  {Sajjad~Athar}}, \ and\ \bibinfo {author} {\bibfnamefont {S.~K.}\
  \bibnamefont {Singh}},\ }\href {\doibase 10.1140/epja/i2018-12534-2}
  {\bibfield  {journal} {\bibinfo  {journal} {Eur. Phys. J.}\ }\textbf
  {\bibinfo {volume} {A54}},\ \bibinfo {pages} {95} (\bibinfo {year}
  {2018}{\natexlab{a}})},\ \Eprint {http://arxiv.org/abs/1802.04469}
  {arXiv:1802.04469 [hep-ph]} \BibitemShut {NoStop}%
\bibitem [{\citenamefont {Fatima}\ \emph
  {et~al.}(2018{\natexlab{b}})\citenamefont {Fatima}, \citenamefont
  {Sajjad~Athar},\ and\ \citenamefont {Singh}}]{Fatima:2018tzs}%
  \BibitemOpen
  \bibfield  {author} {\bibinfo {author} {\bibfnamefont {A.}~\bibnamefont
  {Fatima}}, \bibinfo {author} {\bibfnamefont {M.}~\bibnamefont
  {Sajjad~Athar}}, \ and\ \bibinfo {author} {\bibfnamefont {S.~K.}\
  \bibnamefont {Singh}},\ }\href {\doibase 10.1103/PhysRevD.98.033005}
  {\bibfield  {journal} {\bibinfo  {journal} {Phys. Rev.}\ }\textbf {\bibinfo
  {volume} {D98}},\ \bibinfo {pages} {033005} (\bibinfo {year}
  {2018}{\natexlab{b}})},\ \Eprint {http://arxiv.org/abs/1806.08597}
  {arXiv:1806.08597 [hep-ph]} \BibitemShut {NoStop}%
\bibitem [{\citenamefont {Graczyk}\ and\ \citenamefont
  {Kowal}(2017)}]{Graczyk:2017ngi}%
  \BibitemOpen
  \bibfield  {author} {\bibinfo {author} {\bibfnamefont {K.~M.}\ \bibnamefont
  {Graczyk}}\ and\ \bibinfo {author} {\bibfnamefont {B.~E.}\ \bibnamefont
  {Kowal}},\ }\bibfield  {booktitle} {\emph {\bibinfo {booktitle}
  {{Proceedings, 41st International Conference of Theoretical Physics: Matter
  to the Deepest: Kroczyce, Poland, September 4-8, 2017}}},\ }\href {\doibase
  10.5506/APhysPolB.48.2219} {\bibfield  {journal} {\bibinfo  {journal} {Acta
  Phys. Polon.}\ }\textbf {\bibinfo {volume} {B48}},\ \bibinfo {pages} {2219}
  (\bibinfo {year} {2017})}\BibitemShut {NoStop}%
\bibitem [{\citenamefont {Dombey}(1969)}]{Dombey:1969wk}%
  \BibitemOpen
  \bibfield  {author} {\bibinfo {author} {\bibfnamefont {N.}~\bibnamefont
  {Dombey}},\ }\href {\doibase 10.1103/RevModPhys.41.236} {\bibfield  {journal}
  {\bibinfo  {journal} {Rev. Mod. Phys.}\ }\textbf {\bibinfo {volume} {41}},\
  \bibinfo {pages} {236} (\bibinfo {year} {1969})}\BibitemShut {NoStop}%
\bibitem [{\citenamefont {Donnelly}\ and\ \citenamefont
  {Raskin}(1986)}]{Donnelly:1985ry}%
  \BibitemOpen
  \bibfield  {author} {\bibinfo {author} {\bibfnamefont {T.~W.}\ \bibnamefont
  {Donnelly}}\ and\ \bibinfo {author} {\bibfnamefont {A.~S.}\ \bibnamefont
  {Raskin}},\ }\href {\doibase 10.1016/0003-4916(86)90173-9} {\bibfield
  {journal} {\bibinfo  {journal} {Annals Phys.}\ }\textbf {\bibinfo {volume}
  {169}},\ \bibinfo {pages} {247} (\bibinfo {year} {1986})}\BibitemShut
  {NoStop}%
\bibitem [{\citenamefont {Alguard}\ \emph {et~al.}(1976)\citenamefont {Alguard}
  \emph {et~al.}}]{Alguard:1976bk}%
  \BibitemOpen
  \bibfield  {author} {\bibinfo {author} {\bibfnamefont {M.~J.}\ \bibnamefont
  {Alguard}} \emph {et~al.},\ }\bibfield  {booktitle} {\emph {\bibinfo
  {booktitle} {{Internal spin structure of the nucleon. Proceedings, Symposium,
  SMC Meeting, New Haven, USA, January 5-6, 1994}}},\ }\href {\doibase
  10.1103/PhysRevLett.37.1258} {\bibfield  {journal} {\bibinfo  {journal}
  {Phys. Rev. Lett.}\ }\textbf {\bibinfo {volume} {37}},\ \bibinfo {pages}
  {1258} (\bibinfo {year} {1976})},\ \bibinfo {note} {[,285(1976)]}\BibitemShut
  {NoStop}%
\bibitem [{\citenamefont {Hernandez}\ \emph
  {et~al.}(2007{\natexlab{b}})\citenamefont {Hernandez}, \citenamefont
  {Nieves},\ and\ \citenamefont {Valverde}}]{Hernandez:2007qq}%
  \BibitemOpen
  \bibfield  {author} {\bibinfo {author} {\bibfnamefont {E.}~\bibnamefont
  {Hernandez}}, \bibinfo {author} {\bibfnamefont {J.}~\bibnamefont {Nieves}}, \
  and\ \bibinfo {author} {\bibfnamefont {M.}~\bibnamefont {Valverde}},\ }\href
  {\doibase 10.1103/PhysRevD.76.033005} {\bibfield  {journal} {\bibinfo
  {journal} {Phys. Rev.}\ }\textbf {\bibinfo {volume} {D76}},\ \bibinfo {pages}
  {033005} (\bibinfo {year} {2007}{\natexlab{b}})},\ \Eprint
  {http://arxiv.org/abs/hep-ph/0701149} {arXiv:hep-ph/0701149 [hep-ph]}
  \BibitemShut {NoStop}%
\bibitem [{\citenamefont {Abe}\ \emph {et~al.}(2013)\citenamefont {Abe} \emph
  {et~al.}}]{Abe:2012av}%
  \BibitemOpen
  \bibfield  {author} {\bibinfo {author} {\bibfnamefont {K.}~\bibnamefont
  {Abe}} \emph {et~al.} (\bibinfo {collaboration} {T2K}),\ }\href {\doibase
  10.1103/PhysRevD.87.012001, 10.1103/PhysRevD.87.019902} {\bibfield  {journal}
  {\bibinfo  {journal} {Phys. Rev.}\ }\textbf {\bibinfo {volume} {D87}},\
  \bibinfo {pages} {012001} (\bibinfo {year} {2013})},\ \bibinfo {note}
  {[Addendum: Phys. Rev.D87,no.1,019902(2013)]},\ \Eprint
  {http://arxiv.org/abs/1211.0469} {arXiv:1211.0469 [hep-ex]} \BibitemShut
  {NoStop}%
\bibitem [{\citenamefont {Vermaseren}(2000)}]{Vermaseren:2000nd}%
  \BibitemOpen
  \bibfield  {author} {\bibinfo {author} {\bibfnamefont {J.~A.~M.}\
  \bibnamefont {Vermaseren}},\ }\href@noop {} {\  (\bibinfo {year} {2000})},\
  \Eprint {http://arxiv.org/abs/math-ph/0010025} {arXiv:math-ph/0010025
  [math-ph]} \BibitemShut {NoStop}%
\bibitem [{wcs()}]{wcss}%
  \BibitemOpen
  \href@noop {} {}\bibinfo {howpublished}
  {\url{http://www.wcss.wroc.pl}}\BibitemShut {NoStop}%
\end{thebibliography}%

\end{document}